\documentclass[a4paper,11pt]{article}
\pdfoutput=1 % if your are submitting a pdflatex (i.e. if you have
\usepackage{jinstpub} % for details on the use of the package, please
\usepackage[nooneline]{subfigure}
\subfiguretopcaptrue

\usepackage{pdflscape}

\usepackage{lineno}
\newcommand\Bezier{B\'{e}zier}
\newcommand\tmax{$\theta_\mathrm{max}$}

\title{Prototyping Hexagonal Light Concentrators Using High-Reflectance Specular Films for the Large-Sized Telescopes of the Cherenkov Telescope Array}

\author[a,1]{Akira~Okumura,\note{Corresponding author.}}
\author[b]{Dang~Viet~Tan,}
\author[b]{Sakiya~Ono,}
\author[b]{Syunya~Tanaka,}
\author[c]{Masaaki~Hayashida,}
\author[d]{Jim~Hinton,}
\author[b]{Hideaki~Katagiri,}
\author[e]{Koji~Noda,}
\author[c,f]{Masahiro~Teshima,}
\author[g]{Tokonatsu~Yamamoto,}
\author[b]{Tatsuo~Yoshida,}

\affiliation[a]{Institute for Space--Earth Environmental Research, Nagoya University,\\Furo-cho, Chikusa-ku, Nagoya, Aichi 464--8601, Japan}
\affiliation[b]{College of Science, Ibaraki University, 2--1--1, Bunkyo, Mito, Ibaraki 310--8512, Japan}
\affiliation[c]{Institute for Cosmic Ray Research, The University of Tokyo, Kashiwa, Chiba 277--8582, Japan}
\affiliation[d]{Max-Planck-Institut f\"{u}r Kernphysik, P.O. Box 103980, D 69029 Heidelberg, Germany}
\affiliation[e]{Institut de Fisica d'Altes Energies (IFAE), The Barcelona Institute of Science and Technology, Campus UAB, 08193 Bellaterra (Barcelona), Spain}
\affiliation[f]{Max-Planck-Institut f\"{u}r Physik, F\"{o}hringer Ring 6, D 80805 M\"unchen, Germany}
\affiliation[g]{Department of Physics, Konan University, Kobe, Hyogo 658--8501, Japan}

\emailAdd{oxon@mac.com}

\abstract{We have developed a prototype hexagonal light concentrator for the Large-Sized Telescopes of the Cherenkov Telescope Array. To maximize the photodetection efficiency of the focal-plane camera pixels for atmospheric Cherenkov photons and to lower the energy threshold, a specular film with a very high reflectance of 92--99\% has been developed to cover the inner surfaces of the light concentrators. The prototype has a relative anode sensitivity (which can be roughly regarded as collection efficiency) of about 95 to 105\% at the most important angles of incidence. The design, simulation, production procedure, and performance measurements of the light-concentrator prototype are reported.}

\keywords{Photon detectors for UV, visible and IR photons (vacuum); Cherenkov detectors; Gamma telescopes; Large detector systems for particle and astroparticle physics; Instrument optimisation; Mirror coating; Optics}

\begin{document}
\maketitle
\flushbottom

\section{Introduction}
\label{sec:Introduction}

The photodetection efficiency of the focal-plane cameras of imaging atmospheric Cherenkov telescopes (IACTs) is a key factor that determines the energy threshold for celestial gamma-ray observations in the very-high-energy (VHE, ${\sim}100$~GeV to ${\sim}100$~TeV) regime. This is because most of the photons detected by individual camera pixels originate from the night sky background (NSB), and thus discriminating the faint and transient (${\sim}5$~ns) Cherenkov radiation yielded from VHE gamma-ray cascades from Poisson fluctuation of the NSB requires high quantum efficiency (QE) and collection efficiency of the camera pixels in the waveband of 300--600 nm.

Enlarging the mirror area of an IACT is a straightforward approach to improve the signal-to-noise ratio for low-energy gamma rays. For example, the MAGIC II telescopes have two parabolic reflectors, each with a diameter of 17~m, and the H.E.S.S. II telescope has a diameter of 28~m, achieving threshold trigger energies of ${\sim}50$~GeV and ${\sim}20$~GeV, respectively \cite{Aleksic:2016:The-major-upgrade-of-the-MAGIC-telescopes-Part-II:,Naurois:2017:H.E.S.S.-II---Gamma-ray-astronomy-from-20-GeV-to-h}. It is, however, known from an internal CTA study that the construction cost of a large IACT increases roughly by a power of 1.35 as the mirror area increases \cite{Teshima:2011:Design-Study-of-a-CTA-Large-Size-Telescope-LST}; thus, maximizing the collection efficiency of camera pixels using a less expensive technology to produce light concentrators is a cost-effective way to build a number of IACTs.

The focal-plane cameras of IACTs are covered by UV-sensitive photodetector pixels on which hexagonal light concentrators are usually attached to reduce the dead area between photodetectors \cite{Huber:2011:Solid-light-concentrators-for-small-sized-photosen,Kabuki:2003:Development-of-an-atmospheric-Cherenkov-,Bernlohr:2003:The-optical-system-of-the-H.E.S.S.-imagi,Radu:2000:Design-studies-for-nonimaging-light-conc,Pare:2002:CELESTE:-an-atmospheric-Cherenkov-telesc}. Compound parabolic concentrators (CPC or so-called Winston cones) \cite{Winston:1970:Light-Collection-within-the-Fr} with hexagonal entrance apertures have been widely used in IACTs for this purpose because they can selectively guide photons coming from the telescope-mirror direction to the sensitive area of photodetectors and because those from directions outside the mirror are rejected efficiently at the same time.

The use of Winston cones in IACTs was first proposed for the CAT experiment in 1994 \cite{Punch:1994:Winston-Cones-for-Light-Collection-and-Albedo-Prot}. Since then, a few new technologies, such as direct aluminum coating on plastic cones \cite{Bernlohr:2003:The-optical-system-of-the-H.E.S.S.-imagi,Aguilar:2015:Design-optimization-and-characterization-of-the-li,Henault:2013:Design-of-light-concentrators-for-Cherenkov-telesc} and solid high-UV-transmittance cones utilizing total internal reflection \cite{Oser:2001:High-Energy-Gamma-Ray-Observations-of-the-Crab-Neb,Pare:2002:CELESTE:-an-atmospheric-Cherenkov-telesc,Huber:2011:Solid-light-concentrators-for-small-sized-photosen}, have been developed. The inner-surface profiles of hexagonal light concentrators have also been revisited \cite{Okumura:2012:Optimization-of-the-collection-efficiency-of-a-hex} because the original Winston-cone profile comprising an inclined parabola is optimal only for two-dimensional or rotationally symmetric optical systems.

After the great success of the first pioneering IACT, namely the 10-m Whipple telescope \cite{Weekes:1989:Observation-of-TeV-Gamma-Rays-from-the-C}, and its current-generation successors (the H.E.S.S., MAGIC, and VERITAS telescopes), the Cherenkov Telescope Array (CTA) was proposed as a next-generation ground-based gamma-ray observatory by an international consortium \cite{Actis:2011:Design-concepts-for-the-Cherenkov-Telesc,Acharya:2013:Introducing-the-CTA-concept,CTA-WWW}. The main purposes of the CTA are to significantly improve the gamma-ray-detection sensitivity by an order of magnitude compared to the three telescope arrays currently operating, and to expand the energy coverage, which currently ranges from ${\sim}100$~GeV to ${\sim}20$~TeV, to 20~GeV to over $300$~TeV. These goals will be achieved by building a number of telescopes with different mirror diameters in a large area of several square kilometers in the northern and southern hemispheres. The CTA northern site will comprise four Large-Sized Telescopes (LSTs, 20~GeV--200~GeV) with a mirror diameter of 23~m and 15 Medium-Sized Telescopes (MSTs, 12-m diameter, 100~GeV--10~TeV). The southern site will be comprised of four LSTs, 25 MSTs, and 70 Small-Sized Telescopes (SSTs, 4-m diameter, 5--300~TeV).

A total of eight LSTs serve as the low-energy part of the CTA, reducing the energy threshold of the CTA to as low as 20~GeV to sensitively detect low-energy (20--100~GeV) gamma rays from celestial objects such as gamma-ray bursts and high-redshift active galactic nuclei. The largest mirror size among the CTA is employed to detect faint Cherenkov radiation. To further increase the effective photon-collection area of individual LSTs and to achieve this challenging energy threshold, we have developed new techniques to manufacture light concentrators with very high collection efficiencies in a wide waveband. In the present paper, we report the preliminary design of LST light concentrators and the measured performance of the prototype cones.

\section{Light-concentrator Design}
\label{sec:Design}

\subsection{Design Strategy}
\label{sec:DesignStrategy}

\begin{figure}
  \centering
    \subfigure[]{%
      \label{fig:LST_CG}%
      \includegraphics[width=.34\textwidth,clip]{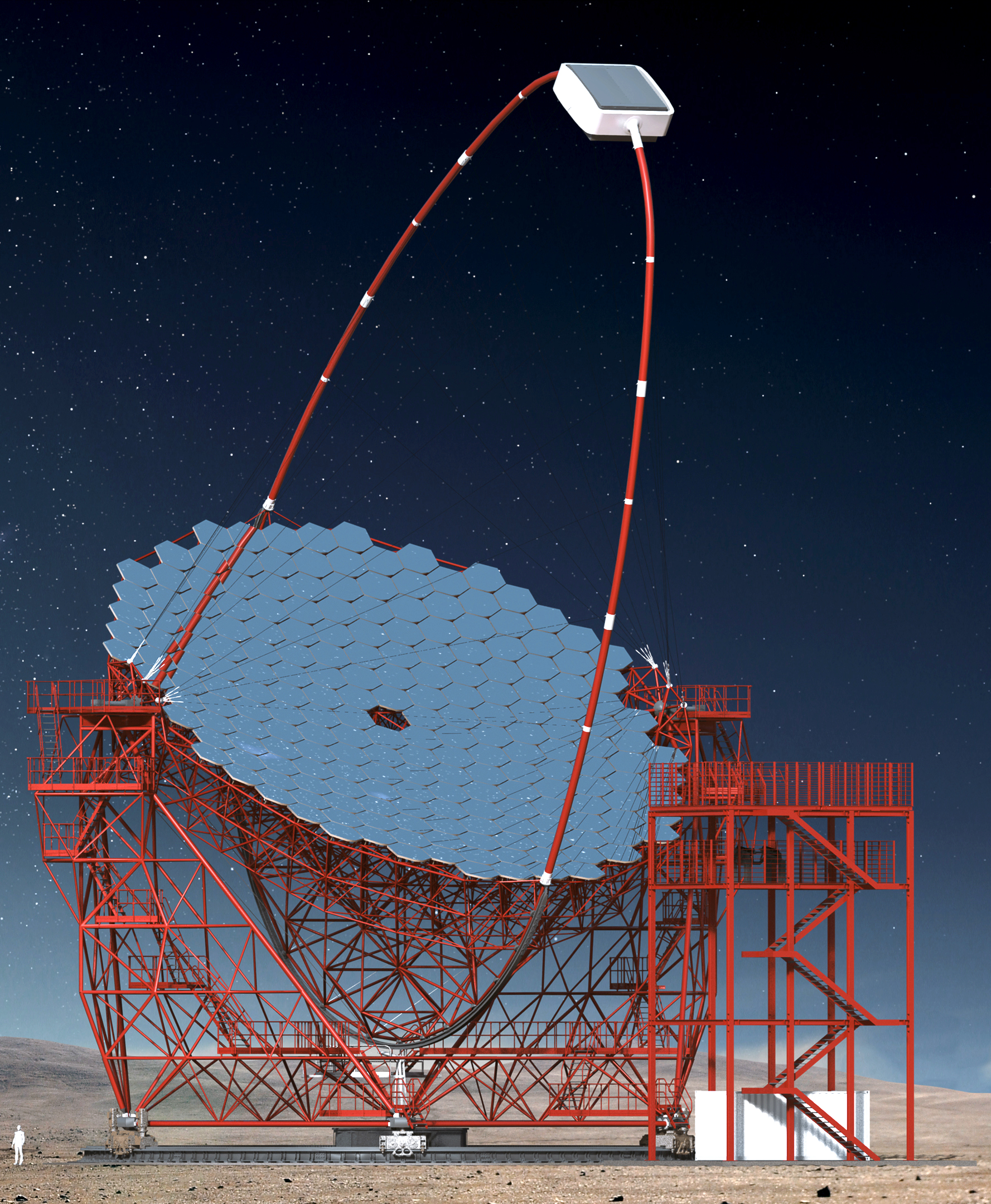}%
    }%
    \subfigure[]{%
      \label{fig:LST_ROBAST}%
      \includegraphics[width=.66\textwidth,clip]{fig/LST.pdf}%
    }
    \caption{(a) Conceptual image of an LST (credit: G. P\'{e}rez, IAC, SMM). (b) \texttt{ROBAST} \cite{Okumura:2016:ROBAST:-Development-of-a-ROOT-based-ray-tracing-li} simulation of a simplified LST optical system comprising 200 segmented mirrors and a focal-plane camera, where the telescope frames are not simulated. The layout and the number of mirror facets are slightly different from those in the final LST design. Blue lines show simulated photon tracks coming from a field angle of $2.15^\circ$.}
    \label{fig:LST}%
\end{figure}

The LST mirror has a pseudo-parabolic surface comprising 198 hexagonal mirror facets\footnote{The number of mirror facets assumed in the present paper is 200 because of a historical reason.} with a side-to-side distance of 1.51~m. Cherenkov photons focused by these mirror facets are imaged by a focal-plane camera comprising 1,855 camera pixels. A conceptual image of an LST and an example of its ray-tracing simulation are shown in Figures~\ref{fig:LST_CG} and \ref{fig:LST_ROBAST}, respectively.

\begin{figure}
  \centering
  \includegraphics[width=.6\textwidth,clip]{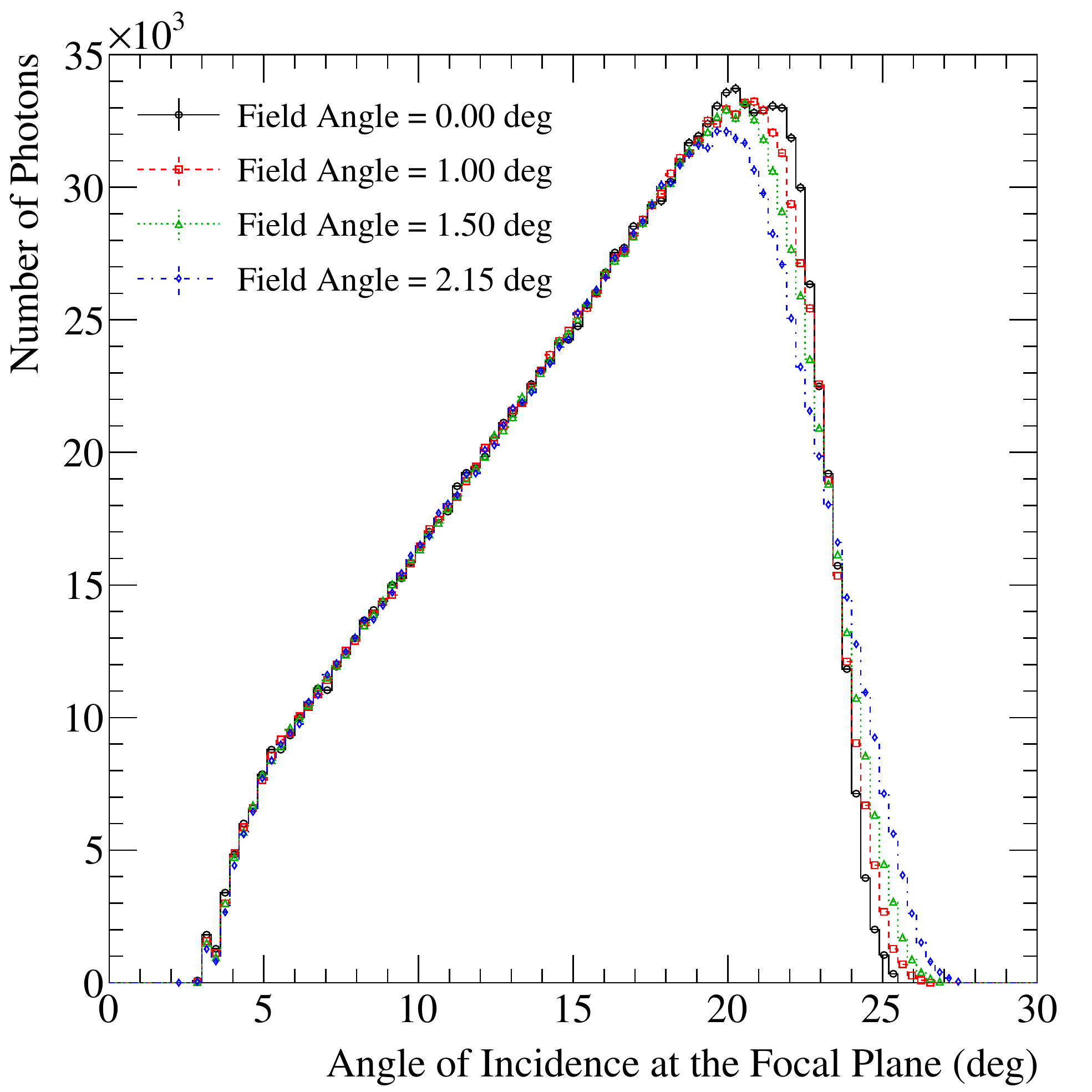}
  \caption{Simulated angular distributions of Cherenkov photons at the LST focal plane for various field angles. Each distribution is averaged over azimuthal angles in the LST field of view. Statistical error bars are not clearly visible because of high statistics.}
  \label{fig:AngDist}
\end{figure}

Cherenkov photons at the focal plane of this parabolic system have a distribution of angle of incidence, $\theta$, ranging from ${\sim}3^\circ$ to ${\sim}25^\circ$ with an approximate weight function $\sin\theta$. Simulated angular distributions for various angles with respect to the optical axis (field angles) are shown in Figure~\ref{fig:AngDist}. LSTs have an effective reflector diameter, $D$, of ${\sim}23$~m and a focal length, $f$, of 28~m, resulting in a maximum angle of incidence of $\tan^{-1} (D/2f) \simeq 22^\circ$. Several outer-mirror facets contribute to larger angles up to ${\sim}27^\circ$. The deficit of 0--$3^\circ$ is due to shadowing by the camera housing and the central hole of the parabolic reflector (see Figure~\ref{fig:LST}), which is reserved for calibration systems.

\begin{figure}
\centering
\includegraphics[width=.8\textwidth,clip]{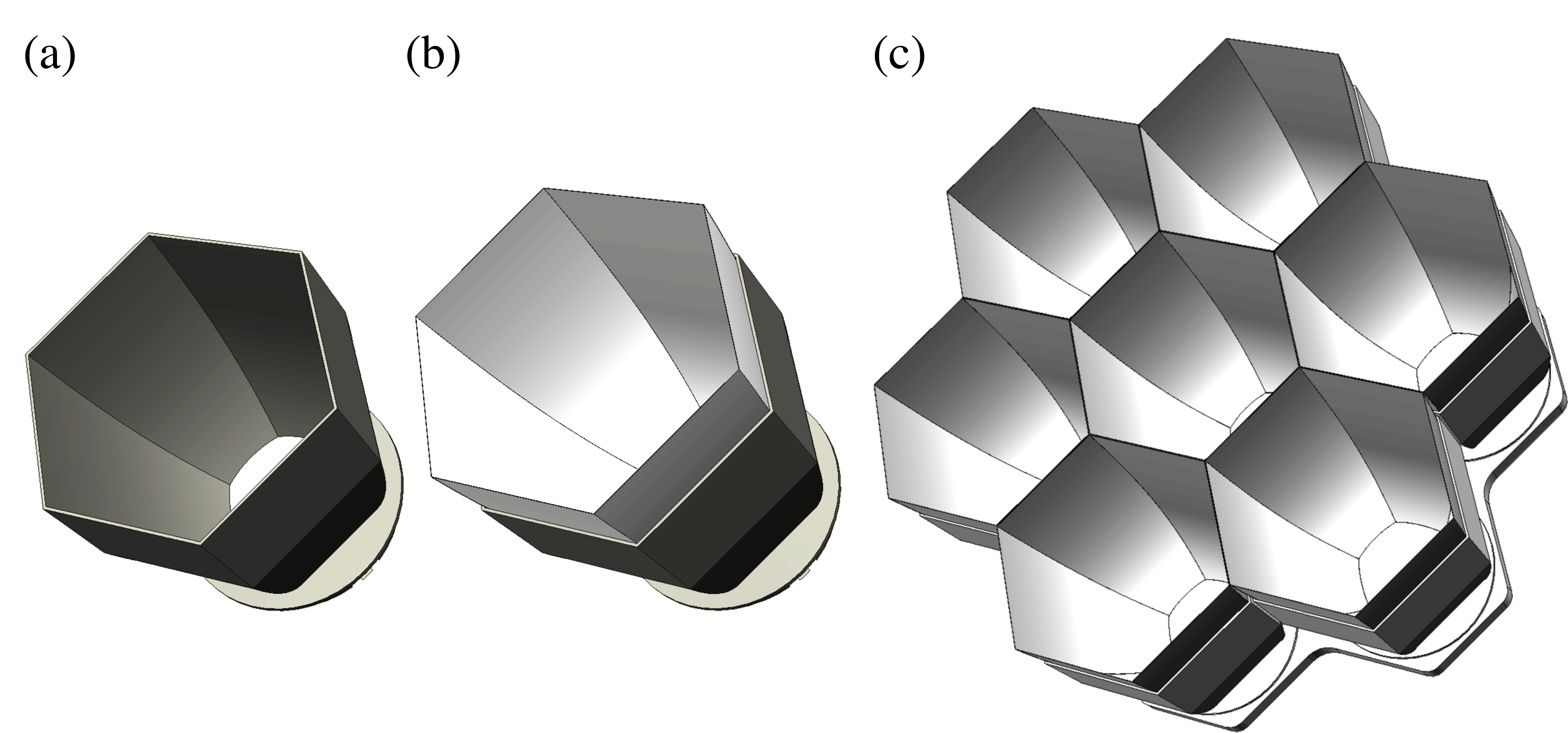}
\caption{(a) 3D CAD model of the base ABS cone of an LST light concentrator prototype. (b) Same as (a) but six specular films are attached to the cone. (c) Light concentrator cluster comprising seven copies of (b) and an interface plate at the bottom. The figures are taken from \cite{Okumura:2015:Prototyping-of-Hexagonal-Light-Concentrators-for-t}.}
\label{fig:CAD}
\end{figure}

LST light concentrators are required to efficiently collect Cherenkov photons using an appropriate inner-surface profile covered by a highly reflective material. We have proposed three new ideas for this purpose. The first is to attach specular films to the inner surface of an injection-molded plastic (acrylonitrile butadiene styrene, ABS) cone. This makes it possible to maximize the entrance areas of individual light concentrators, as illustrated in Figure~\ref{fig:CAD}. The thicknesses of such films are expected to be less than 100~$\mu$m, while it is technically difficult to mold thin plastic edges below ${\sim}500$~$\mu$m. If the upper side of each specular film is kept free (i.e., the plastic-cone height is shorter than the film length, as shown in Figure~\ref{fig:CAD}), the coverage of the entrance area can be $((25\ \mathrm{mm} - 0.1\ \mathrm{mm})/25\ \mathrm{mm})^2 = 99.2\%$ when the side-to-side distance of a single pixel is 50~mm. On the contrary, if we use a plastic cone throughout from the bottom to the top, the coverage is reduced to a maximum of $((25\ \mathrm{mm} - 0.5\ \mathrm{mm})/25\ \mathrm{mm})^2 = 96.0\%$.

\begin{figure}
  \centering
  \subfigure[]{%
    \includegraphics[width=.5\textwidth,clip]{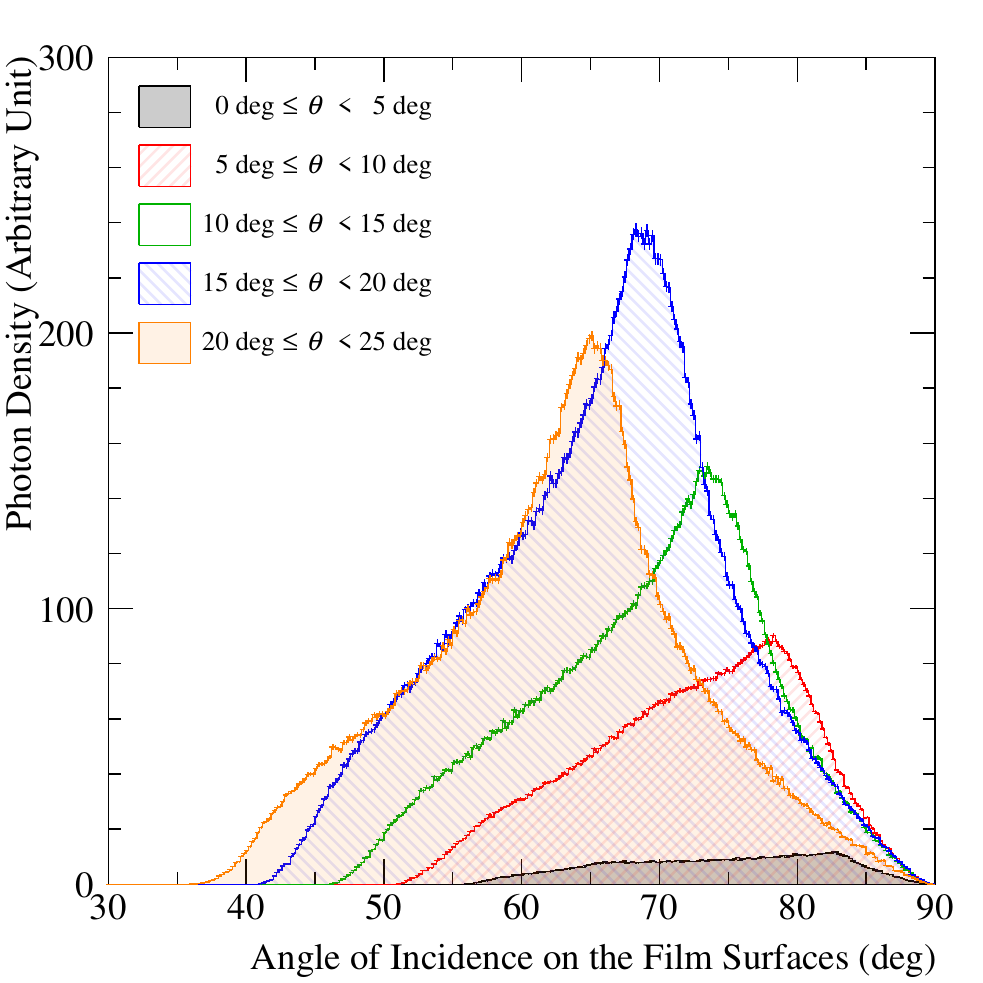}
    \label{fig:ReflectionAngleDist}
  }%
  \subfigure[]{%{
    \includegraphics[width=.5\textwidth,clip]{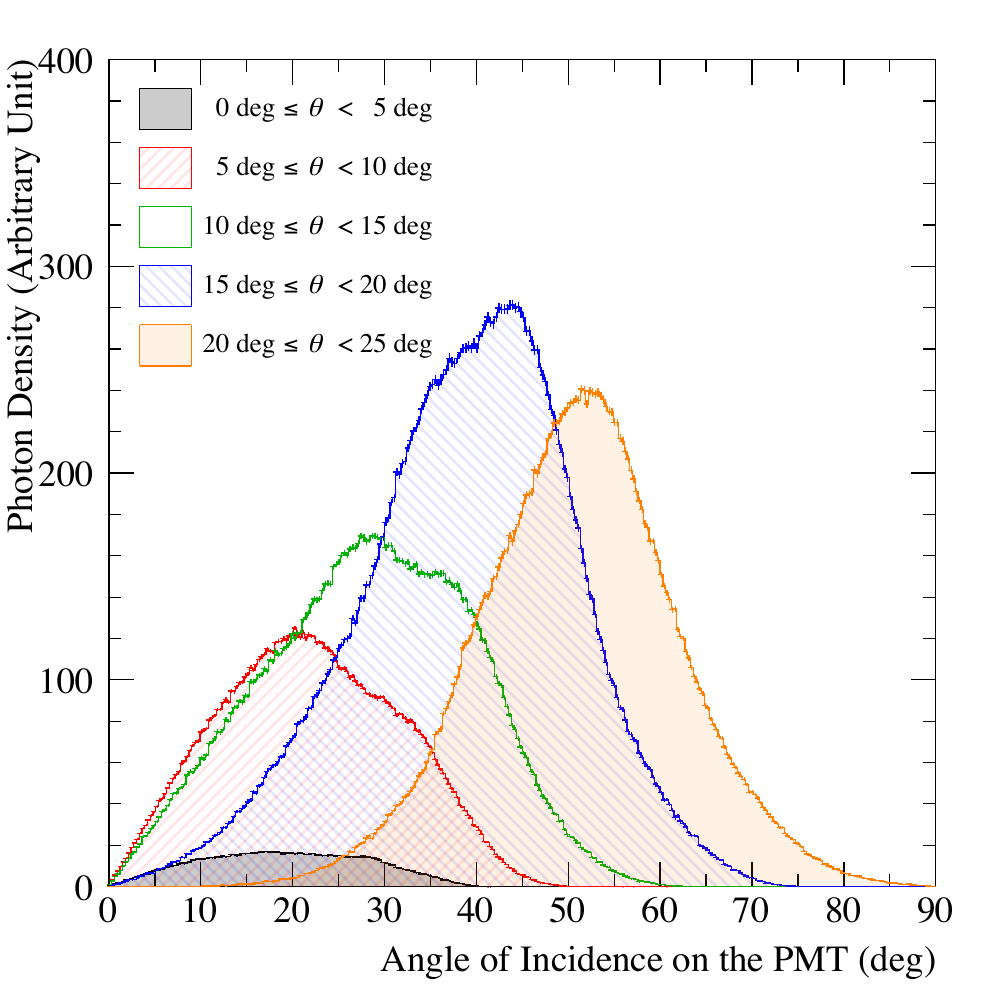}
    \label{fig:PMTAngleDist}
  }%
  \caption{(a) Simulated distribution of angles of incidence on the film surfaces. Only the first reflections of photons that are detected later by the PMT are used. $\theta$ is the angle of incidence at the cone entrance. The simulation was performed for the optimized cone profile described in Section~\ref{sec:Optimization} with a constant reflectance of 95\%. The photon angular distribution at the focal plane for a field angle of 1.50$^\circ$ (Figure~\ref{fig:AngDist}) was assumed, resulting in the small fraction for 0--5$^\circ$. (b) Simulated distribution of angles of incidence on the PMT.}
\end{figure}

\begin{figure}
  \centering
  \subfigure[]{%
    \includegraphics[width=.5\textwidth,clip]{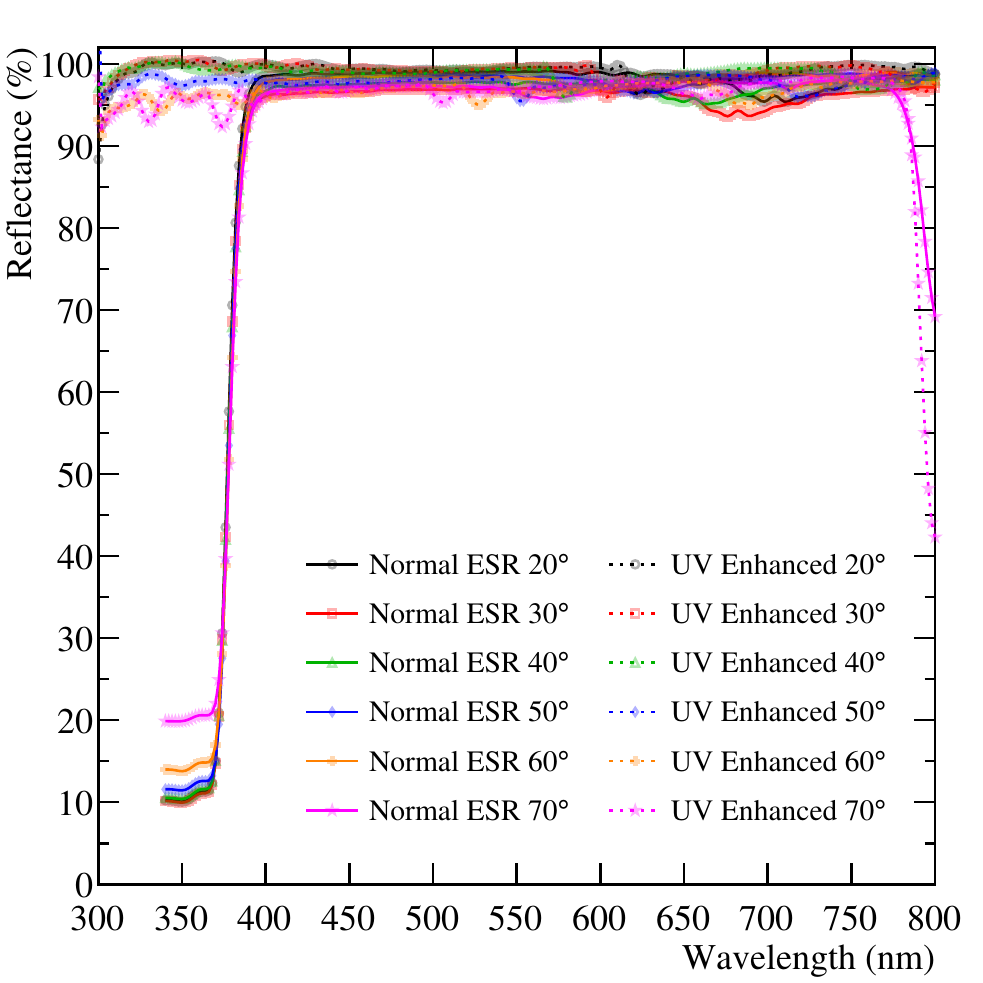}%
    \label{fig:reflectance_full}%
  }%
  \subfigure[]{%
    \includegraphics[width=.5\textwidth,clip]{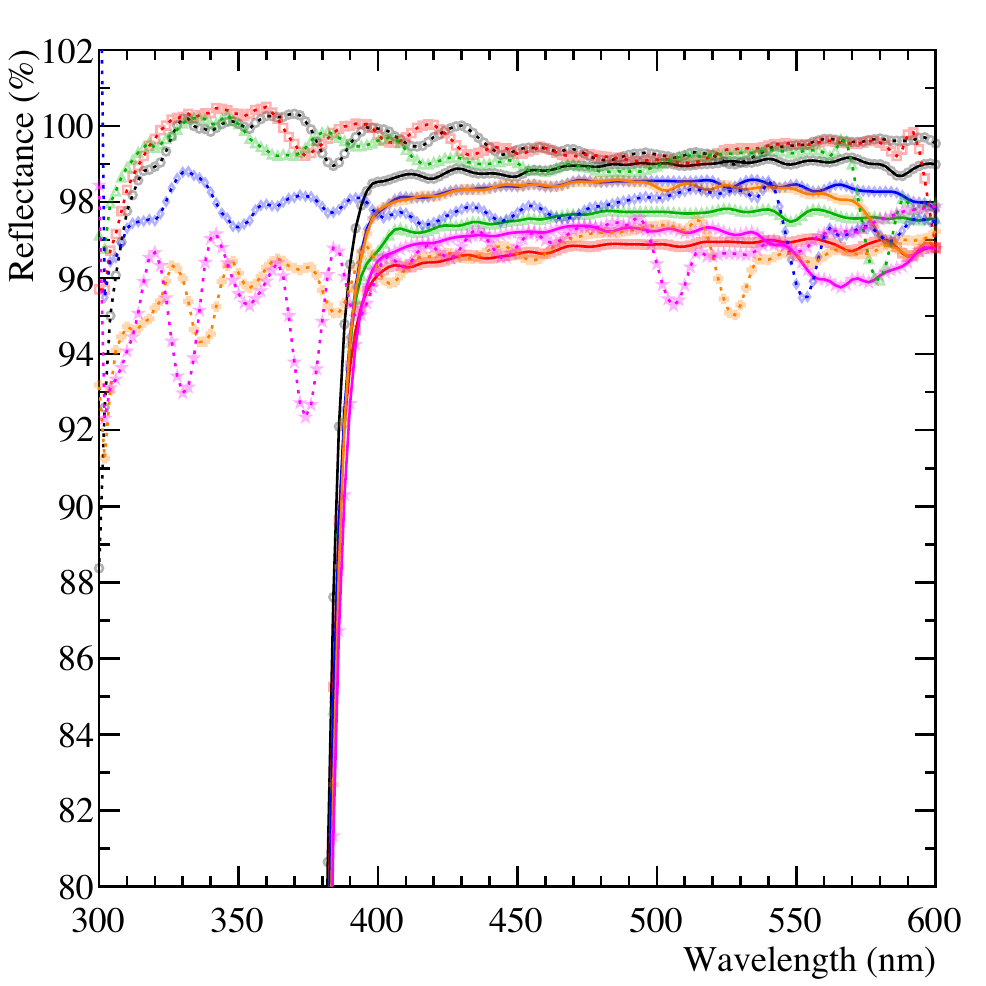}%
    \label{fig:reflectance_zoomin}%
  }
  \caption{(a) Measured reflectance of normal (solid) and UV-enhanced (dashed) ESR films. Because of the measurement limitation of the spectrometer, Hitachi High-Tech U-4100 with an optional attachment, variable angle absolute reflectance accessory (product number 134-0115, guaranteed wavelengths 340 to 2,000~nm), data points below 320~nm, which are not very reliable, have large fluctuations and start dropping around 310~nm regardless of angles of incidence. (b) Same as (a) but only limited axis ranges are shown. The systematic uncertainty of these measurements, which mainly comes from the nonlinearity and noise of the photomultiplier in the spectrometer, is estimated to be ${\sim}1$\%; thus, some data points are higher than 100\%.}
  \label{fig:reflectance}%
\end{figure}

Our second idea is to utilize an existing specular-film product (Vikuiti ESR by 3M, hereafter referred to as ESR) with a thickness of 65~$\mu$m for the inner-reflector films. While the nominal ESR reflectance of 400--800~nm is 98\% or more for various angles of incidence, it is known that the reflectance drops below 400~nm, which is the important waveband for detecting Cherenkov photons. Therefore, an additional multilayer coating (54 layers of Ta$_2$O$_5$ and SiO$_2$) was applied to the ESR surface to enhance the reflectance in the UV range between 300 and 400~nm\footnote{The multilayer coating was done by Bte Bedampfungstechnik GmbH, Germany.}. The multilayer design was optimized for an angle of incidence of $65^\circ$ because those on the film surface range mostly from $50^\circ$ to $80^\circ$ as shown in Figure~\ref{fig:ReflectionAngleDist}. Figure~\ref{fig:reflectance} compares measured reflectance vs. wavelength for various angles of incidence. A normal ESR film exhibits very low reflectance ($<$10\%) below 370~nm due to UV absorption, but the additional multilayer coating pulls up the reflectance to values ranging from about 92\% up to more than 99\% in this band. The reflectance in the range of 300--600~nm is higher than those of direct aluminum and multilayer coating on ABS by 2--10\% at angles ranging from $40^\circ$ to $70^\circ$ (see, for example, \cite{Aguilar:2015:Design-optimization-and-characterization-of-the-li} for simulated reflectance of the direct coating).

The third idea is to use a cubic \Bezier{} curve for the inner-surface profile  because it is known that conventional Winston cones are not optimal for hexagonal light concentrators and because the use of a \Bezier{} curve improves the collection efficiency by ${\sim}2$\% when the entrance and exit apertures are hexagons \cite{Okumura:2012:Optimization-of-the-collection-efficiency-of-a-hex}. As a result, our prototype light concentrators are expected to have higher collection efficiencies compared with the other prototypes for SSTs \cite{Aguilar:2015:Design-optimization-and-characterization-of-the-li} and MSTs \cite{Henault:2013:Design-of-light-concentrators-for-Cherenkov-telesc} (by roughly up to 10\% in total).

\subsection{Optimization of the Surface Profile}
\label{sec:Optimization}

The inner-surface profile of the ABS cones was optimized by following the \Bezier-curve method described in \cite{Okumura:2012:Optimization-of-the-collection-efficiency-of-a-hex}. In addition, we introduced a few assumptions to the ray-tracing simulation to make it more realistic than the simulation performed in \cite{Okumura:2012:Optimization-of-the-collection-efficiency-of-a-hex}.

First, the reflectance of the specular surfaces was assumed to be 95\% for any angle of incidence, while only 90\% and 100\% reflectances were used in \cite{Okumura:2012:Optimization-of-the-collection-efficiency-of-a-hex}. As already shown in Figure~\ref{fig:reflectance}, the measured reflectance in fact surpasses 95\%; however, the surface profile of our prototype was determined before obtaining the coating result, and thus a typical lower limit of 95\% was employed.

\begin{figure}
\centering
\includegraphics[width=.6\textwidth,clip]{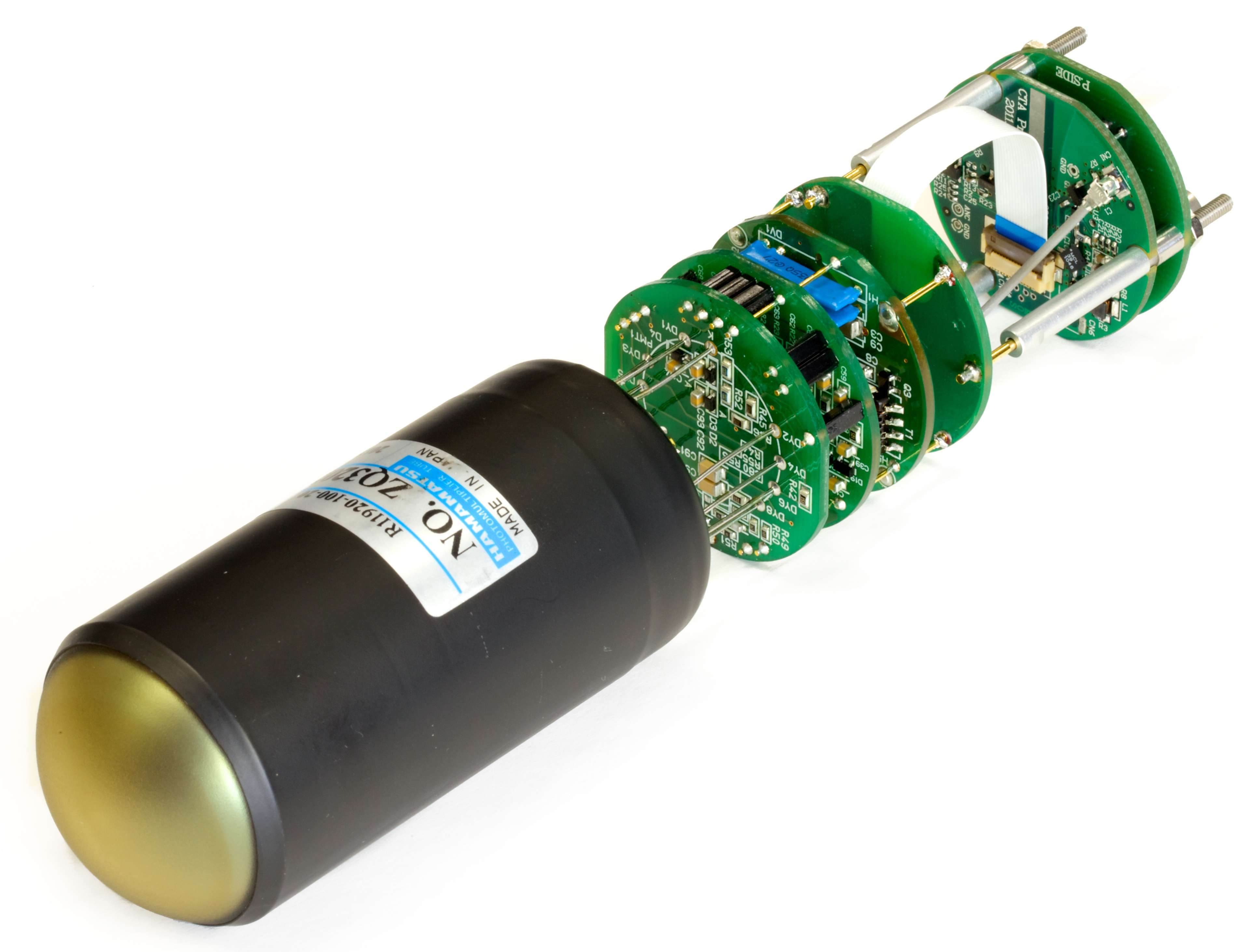}
\caption{Prototype LST PMT module comprising a semispherical PMT (Hamamatsu Photonics R11920-100-20) and preamplifier and high-voltage modules \cite{Inome:2014:Development-of-the-camera-for-the-large-size-teles}.}
\label{fig:PMT}
\end{figure}

The second assumption is that the photodetector attached to individual light concentrators is a photomultiplier tube (PMT) with a semispherical photo cathode (Hamamatsu Photonics R11920-100-20\footnote{See reference \cite{Toyama:2013:Novel-Photo-Multiplier-Tubes-for-the-Cherenkov-Tel} for the characteristics of a similar PMT model R11920-100-05.} or R12992-100-05, radius of curvature of 21~mm, tube diameter of 39.6~mm). The main difference between R11920 and R12992 is the number of dynodes. The former has eight dynodes, while the latter has seven. R11920 is intended for use in the first LST camera to be installed on the island of La Palma, while R12992 will be used for the other seven cameras. In our measurement (to be described in Section~\ref{sec:MeasRes}), R11920 was used. The entrance window of these PMTs has a matte surface to improve the effective QE\footnote{The intrinsic QE of the photocathode does not change; however, it is considered that the photons reflected from the photocathode are trapped inside the window owing to scattering at the matte boundary \cite{Paneque:2003:A-method-to-enhance-the-sensitivity-of-photomultip}.}, as shown in Figure~\ref{fig:PMT}.

\begin{figure}
  \centering
  \subfigure[]{%
    \includegraphics[width=.5\textwidth,clip]{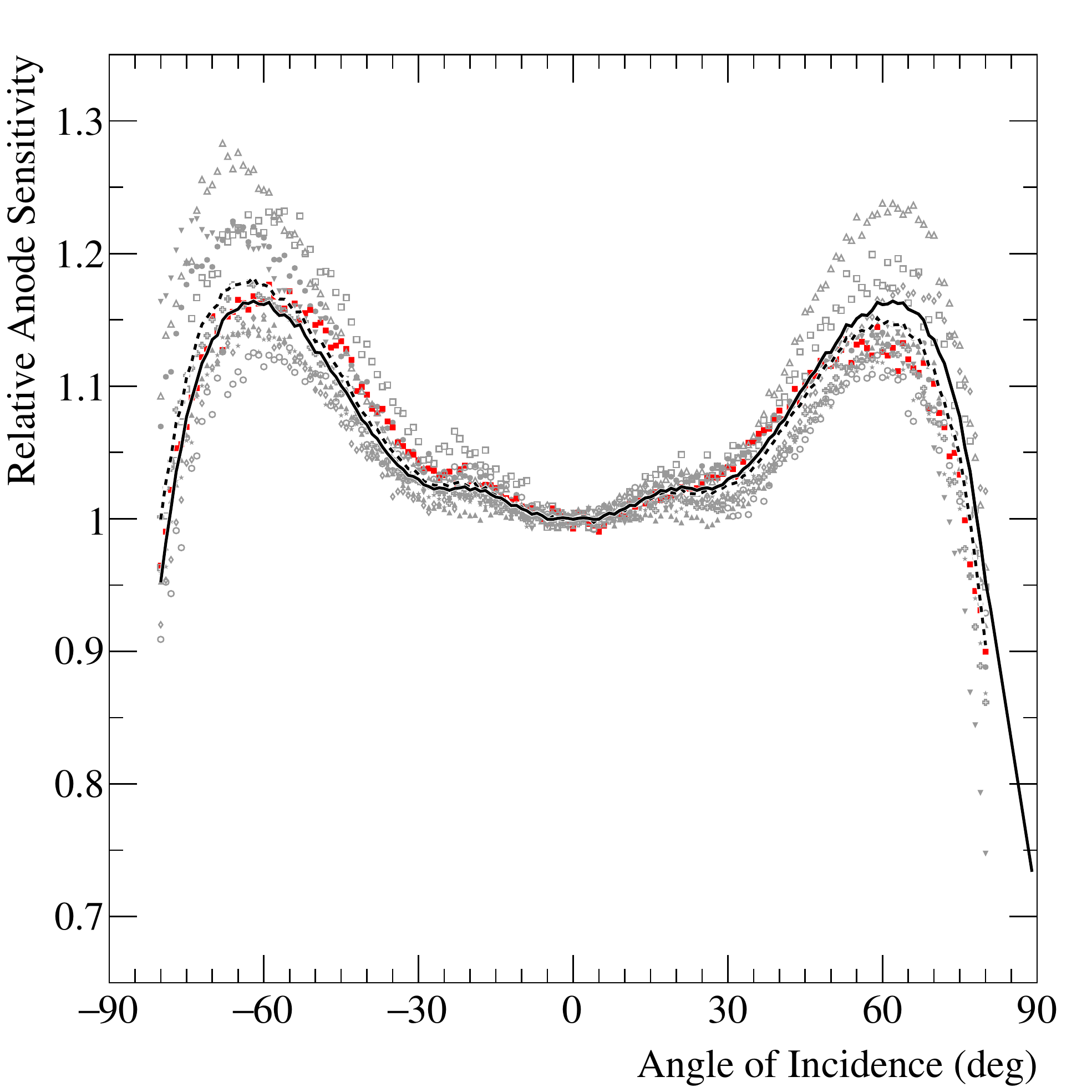}%
    \label{fig:PMT_angle}%
  }%
  \subfigure[]{%
    \includegraphics[width=.5\textwidth,clip]{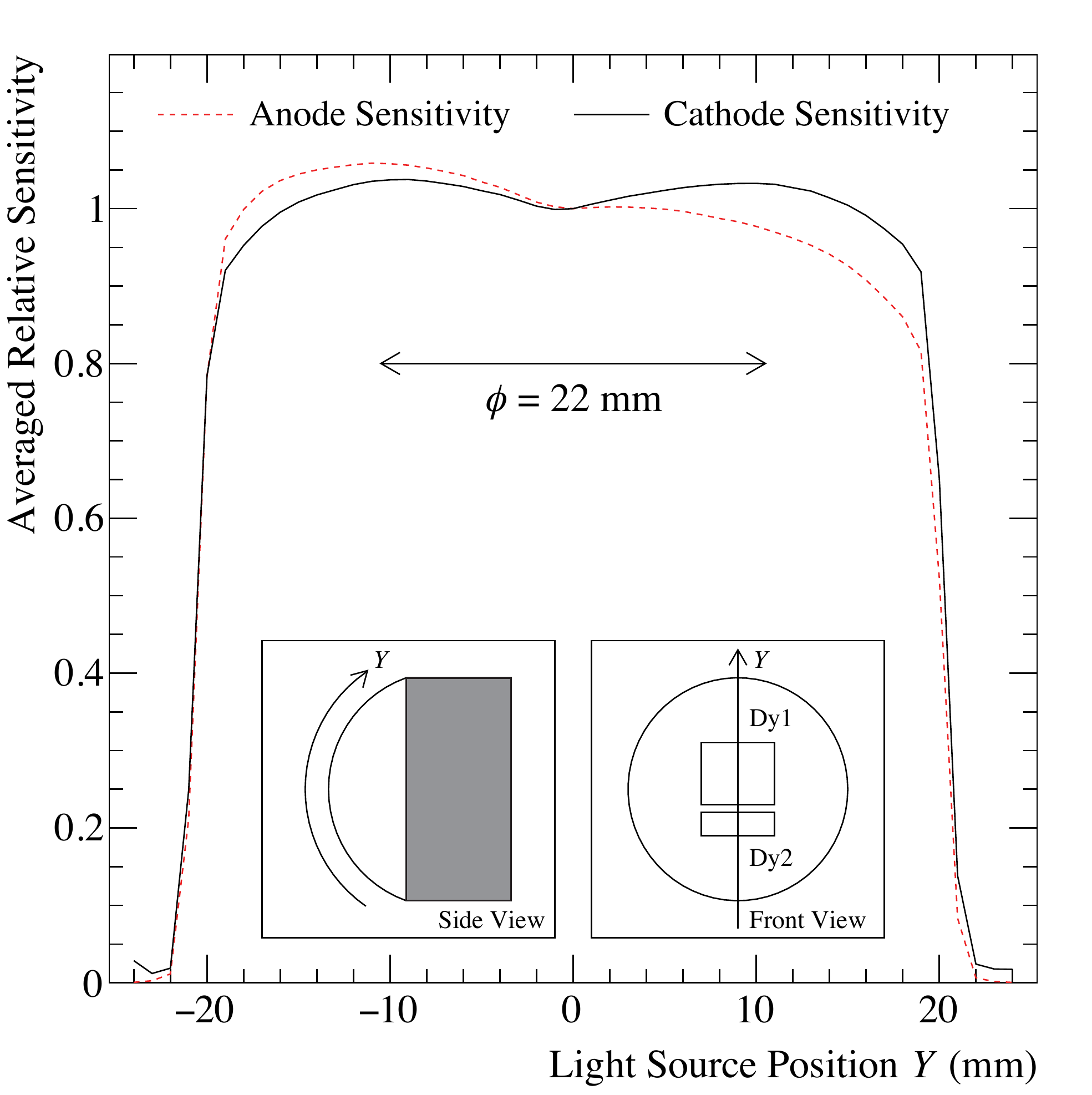}%
    \label{fig:PMT_position}%
  }%
  \caption{(a) Anode sensitivity for 465-nm photons versus the angle of incidence, normalized relative to the vertically incident (i.e., $0^\circ$) photons. The data points show the values measured of eight PMTs (R11920). The values measured for the PMT used in the evaluation of light concentrators are shown with red-filled triangles. The dashed line shows the average of the eight PMTs. The solid line shows the symmetrical average (and extrapolates to 80--$90^\circ$). The systematic uncertainty of data points mainly caused by instability of the light source and the PMT nonlinearity is typically about 1\%. (b) The averaged anode (red dashed) and cathode (black solid) sensitivity of LST PMTs (R12992) versus light source position. Measurements of 90 PMTs were averaged and normalized to the position at $Y = 0$~mm. The insets schematically illustrate a PMT and the definition of the curved $Y$ axis. The systematic uncertainty of the curves here is again about 1\%. The figures are taken from \cite{Okumura:2015:Prototyping-of-Hexagonal-Light-Concentrators-for-t}.
    \label{fig:PMT_response}
  }
\end{figure}

It is known from a previous study \cite{Paneque:2003:A-method-to-enhance-the-sensitivity-of-photomultip} that PMTs with a semispherical entrance window have ``double-crossing effect,'' that is, a photon transmitted through the photocathode may cross the photocathode again. In addition the effective thickness of the photocathode increases as the angle of incidence at the normal to the entrance window increases. As a result, the effective QE increases when the angle of incidence is large. We measured these effect on eight PMTs (R11920) to take it into account in our ray-tracing simulation. In this measurement a mask with a 5~mm-diameter hole was placed at the centers of the entrance windows of the individual PMTs installed on a motorized rotation stage, and a blue light-emitting diode (LED, Nichia NSPB346KS, peak at 465~nm) located 2.4~m away was illuminated toward the hole. Figure~\ref{fig:PMT_angle} compares the measurements for the eight PMTs and their average, in which an obvious increase in anode sensitivity (proportional to the product of QE and the dynode collection efficiency) up to an average of 16\% better for favorable angles of incidence of 60$^\circ$ in comparison to that at $0^\circ$ is observed.

Nonuniformity of the photocathode was also measured by Hamamatsu Photonics for 90 PMTs (R12992), as shown in Figure~\ref{fig:PMT_position}. The positional dependence is smaller than the angular dependence but it is also taken into account in the ray-tracing simulation.

\texttt{ROOT-based simulator for ray tracing} (\texttt{ROBAST}) \cite{Okumura:2016:ROBAST:-Development-of-a-ROOT-based-ray-tracing-li} was used to perform ray-tracing simulation for optimization of the cone surface profile, in other words, to maximize the number of Cherenkov photons that PMTs can detect\footnote{While a prototype MST light concentrator has been designed to maximize the signal-to-noise ratio (signal: Cherenkov photons, noise: fluctuation of the NSB) \cite{Henault:2013:Design-of-light-concentrators-for-Cherenkov-telesc}, we have intentionally chosen a different approach to maximize the number of detected Cherenkov photons because Poisson fluctuation is more critical for low-energy photons (${\sim}20$~GeV).}.  As the pixel size of the LST camera is defined as exactly 50~mm (side-to-side), the aperture radius, $\rho_1$, of the light concentrators is a fixed parameter. On the other hand, six parameters were set as free in the optimization: the exit-aperture radius, $\rho_2$; cone height, $L$; and coordinates of the two control points, $\vec{P}_1$ and $\vec{P}_2$, of a cubic \Bezier{} curve (see \cite{Okumura:2012:Optimization-of-the-collection-efficiency-of-a-hex} for the details).

Taking into account the thickness of the ESR films (65~$\mu$m) and its small distortion caused by misalignment and tolerance of individual cones, $\rho_1$ was set to be 24.9~mm. Nine different combinations of $\rho_2$ (10.0, 10.5, and 11.0~mm) and $L$ (65.0, 66.5, and 68.0~mm) were simulated in a four-dimensional parameter space of $\vec{P}_1$ and $\vec{P}_2$. The best parameter set that most efficiently detects photons entering with the angular distribution shown in Figure~\ref{fig:AngDist} was chosen. $\rho_2$ was set to 11.0~mm or smaller to suppress non-Cherenkov photons coming from the outside of the parabolic reflector. In addition, owing to the geometrical size of ESR films that are commercially available for small retail customers, $L$ was set to 68.0~mm or smaller. The best parameter set maximizing the photodetection efficiency was $L= 68.0$~mm, $\rho_2=11.0$~mm, $\vec{P}_1=(0.55, 0.20)$, and $\vec{P}_2 = (0.90, 0.45)$.

\begin{figure}
\centering
\includegraphics[width=.6\textwidth,clip]{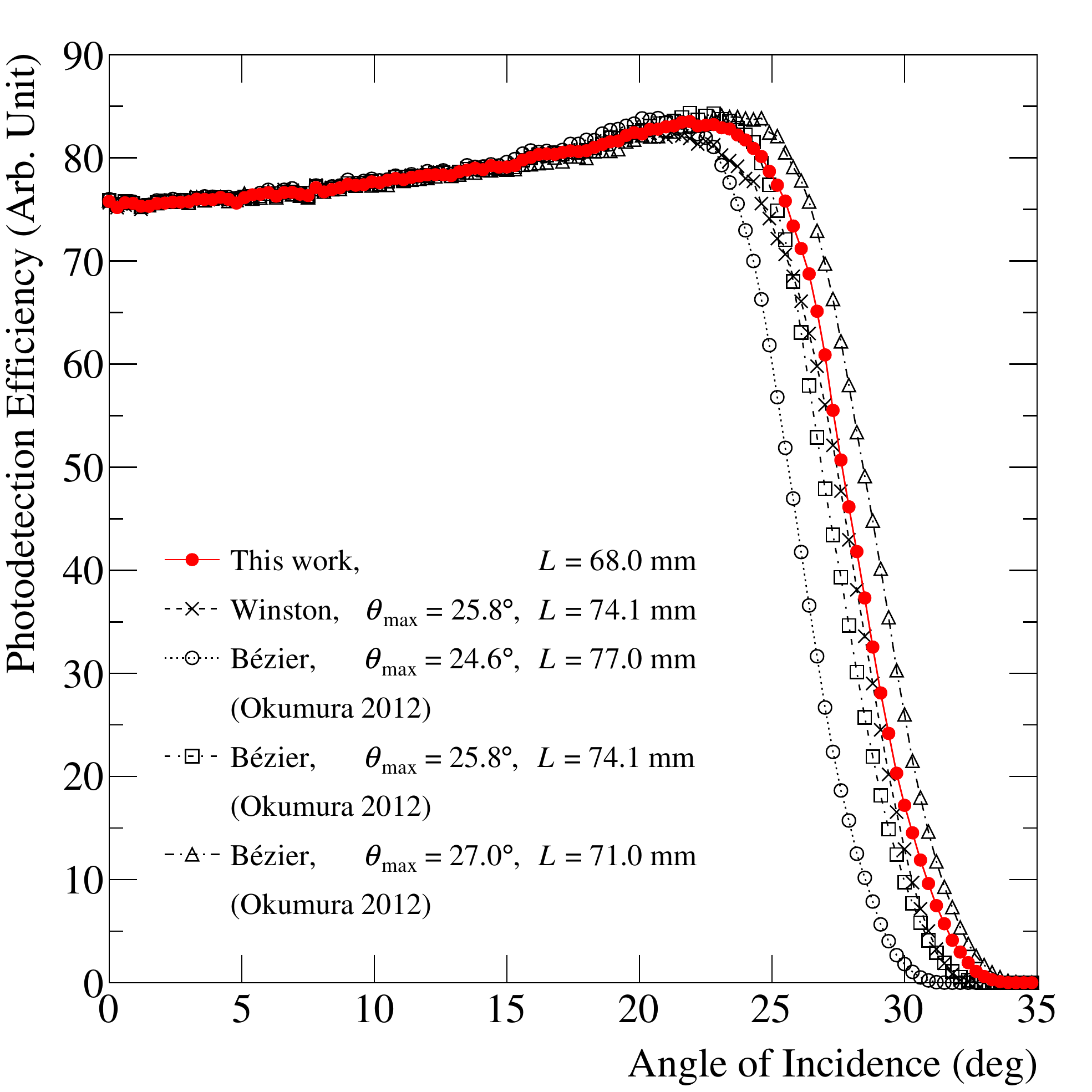}
\caption{Comparison of simulated photodetection efficiencies of the optimized cone profile (filled circles), a conventional Winston cone with a \tmax{} of $25.8^\circ$ (crosses), and \Bezier{} cones with \tmax{} values of $24.6^\circ$ (open circles), $25.8^\circ$ (open squares), and $27.0^\circ$ (open triangles). A hexagonal entrance aperture of 49.8~mm is assumed for all the profiles. $L$ in the legend represents the height of each cone. Note that \tmax{} cannot be defined for the optimized cone profile because the profile parameterization is different from the others.}
\label{fig:cone_simulations}
\end{figure}

Figure~\ref{fig:cone_simulations} compares the simulated relative photodetection efficiencies of five different cone profiles as functions of the angle of incidence. All the curves show a ``shoulder'' structure around $22^\circ$, which exceeds the efficiency at $0^\circ$ by ${\sim}10$\%. This shoulder structure results from the distribution of angles of incidence on the PMT and the angular dependence of the anode sensitivity and, as shown in Figure~\ref{fig:PMTAngleDist} and Figure~\ref{fig:PMT_angle}, respectively. Indeed the assumption of a flat anode sensitivity for all angles of incidence in the simulation removes this structure. Note that the vertical axis is not the collection efficiency of light concentrators, it is instead the photodetection efficiency (i.e., the product of the collection efficiency of a light concentrator, the effective QE of a PMT, and the collection efficiency of the PMT) because the collection efficiency cannot be defined due to the angular dependence of the anode sensitivity.

The optimized cone profile shown in Figure~\ref{fig:cone_simulations} has a few notable features. First, its peak around $23^\circ$ is higher than that of the conventional Winston cone. This is because using a \Bezier-curve profile for a hexagonal light concentrator improves collection efficiency and offers a sharper cutoff, as explained in \cite{Okumura:2012:Optimization-of-the-collection-efficiency-of-a-hex}. Secondly, the optimized profile has a similar efficiency curve to those of the \Bezier{} profiles with \tmax~($\equiv \arcsin(\rho_2/\rho_1)$, also see \cite{Okumura:2012:Optimization-of-the-collection-efficiency-of-a-hex}) of $25.8^\circ$ and $27.0^\circ$, even though a shorter cone length, $L$, of 68~mm is used. This feature is preferable because of the physical-size limitation of available ESR sheets\footnote{For small retail customers, the maximum diagonal length of a sheet is 200~mm.}.

\section{Production}
\label{sec:Production}

\begin{figure}
  \centering
  \subfigure[]{%
    \includegraphics[width=.5\textwidth,clip]{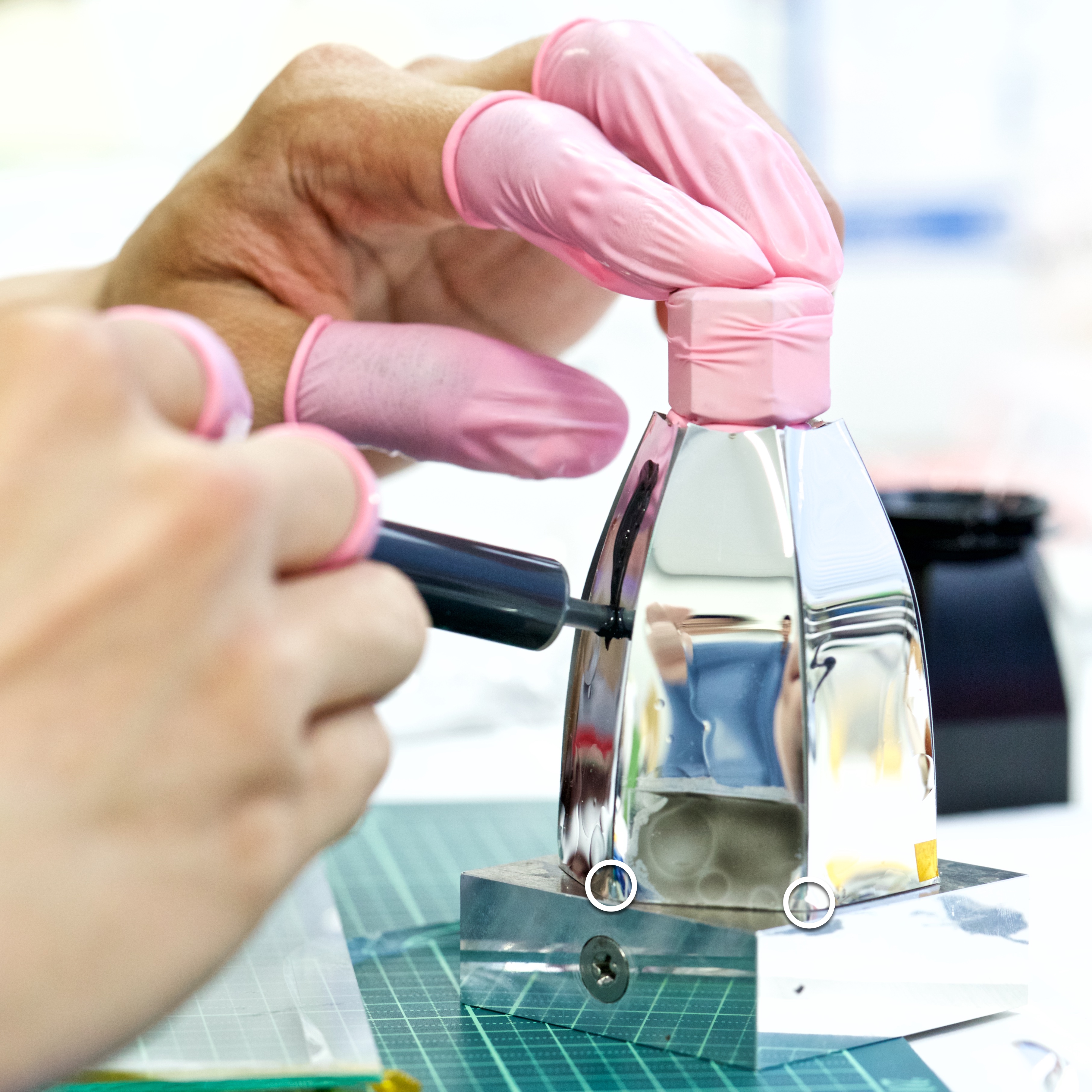}%
    \label{fig:Glueing_with_hands}%
  }%
  \subfigure[]{%
    \includegraphics[width=.5\textwidth,clip]{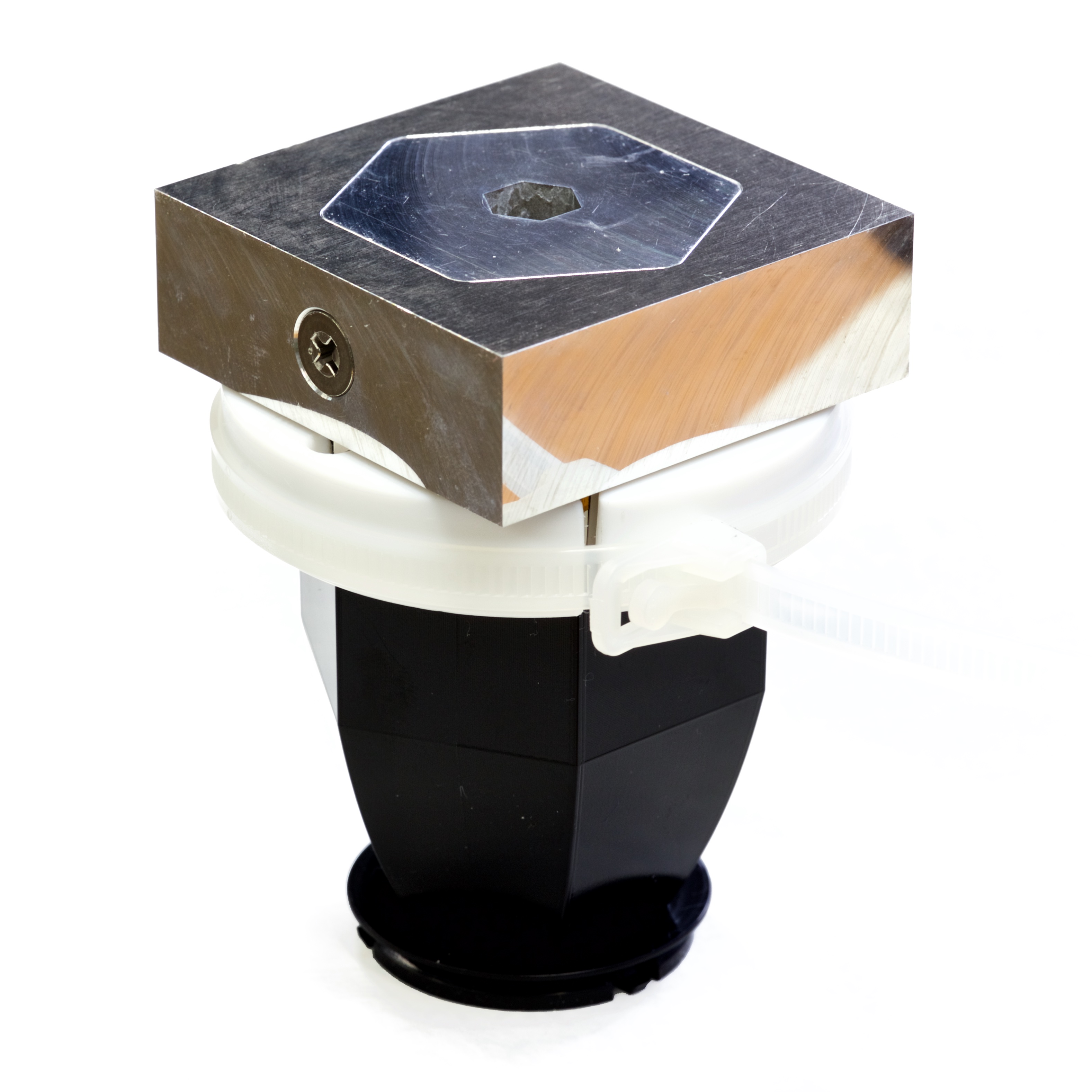}%
    \label{fig:Glueing_with_cable_binder}%
    }
\caption{(a) Photo of the production process at Ibaraki University. Six ESR pieces are placed on a ``male'' block. The white circles indicate the positions which are shared by neighbor pieces. One of such six positions is fixed by thin Kapton tape. (b) Photo of a light concentrator during the gluing process of the ESR films. The aluminum male block shown in (a) works as a weight which presses ESR films on the inner surface of the black ABS cone. Silicone glue is spread between the films and the cone. A white cable binder and Teflon block are used for glueing steel plates to the backside of the ESR films to keep the film flat.}
\label{fig:Glueing}
\end{figure}

The six specular surfaces of a light concentrator are made of six pieces of UV-enhanced ESR. The pieces are cut from a single ESR sheet with dimensions of 76~mm $\times$ 184.9~mm (with a diagonal length of 199.9~mm) using a CO$_2$-laser cutter. These pieces share ${\sim}5$-mm edges with their neighbors such that workers can easily glue the six films together onto a plastic cone at the same time (see the circles in Figure~\ref{fig:Glueing_with_hands}).

\begin{figure}
  \centering
  \subfigure[]{%
    \includegraphics[width=.5\textwidth,clip]{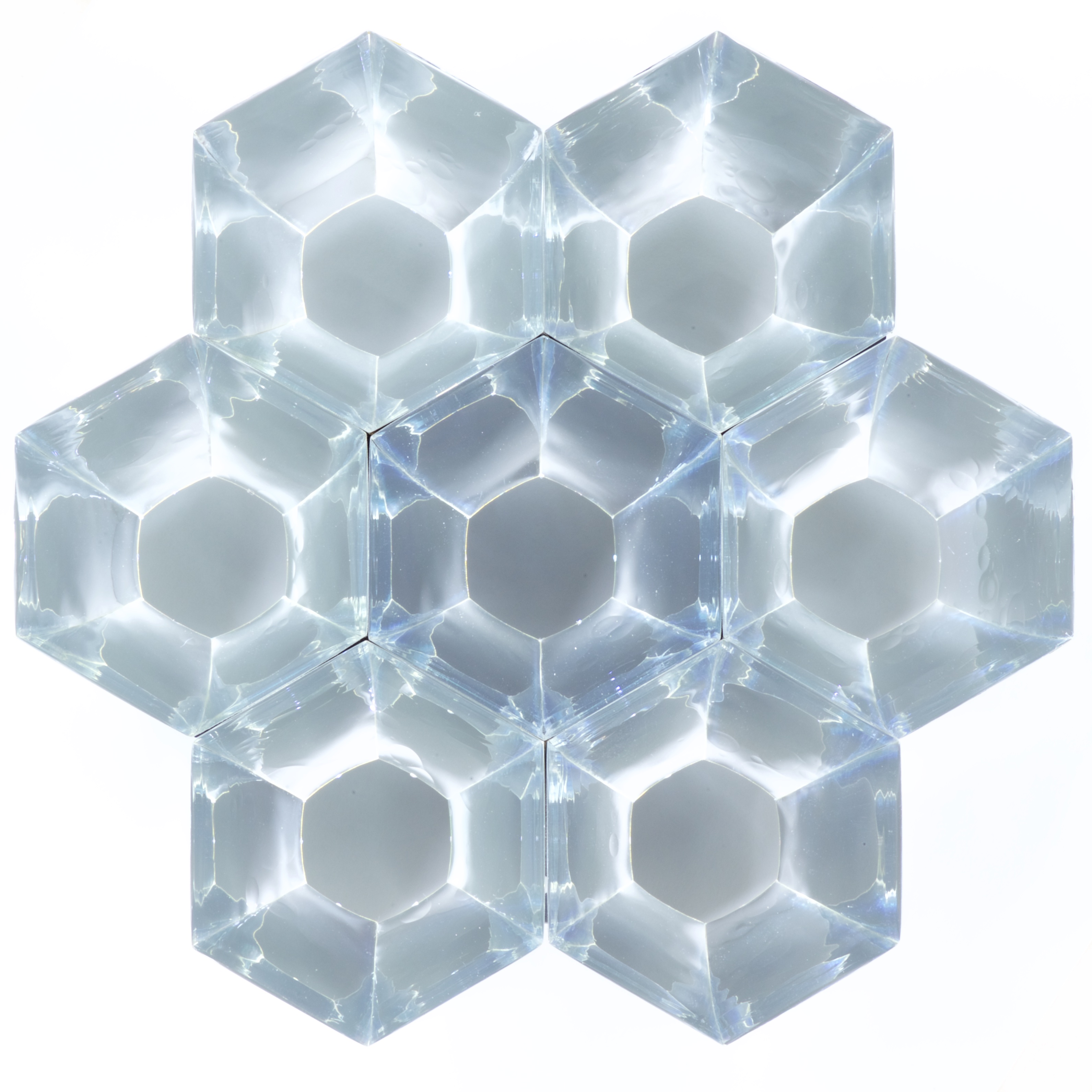}%
    \label{fig:7LCs_top_view}
  }%
  \subfigure[]{%
    \includegraphics[width=.5\textwidth,clip]{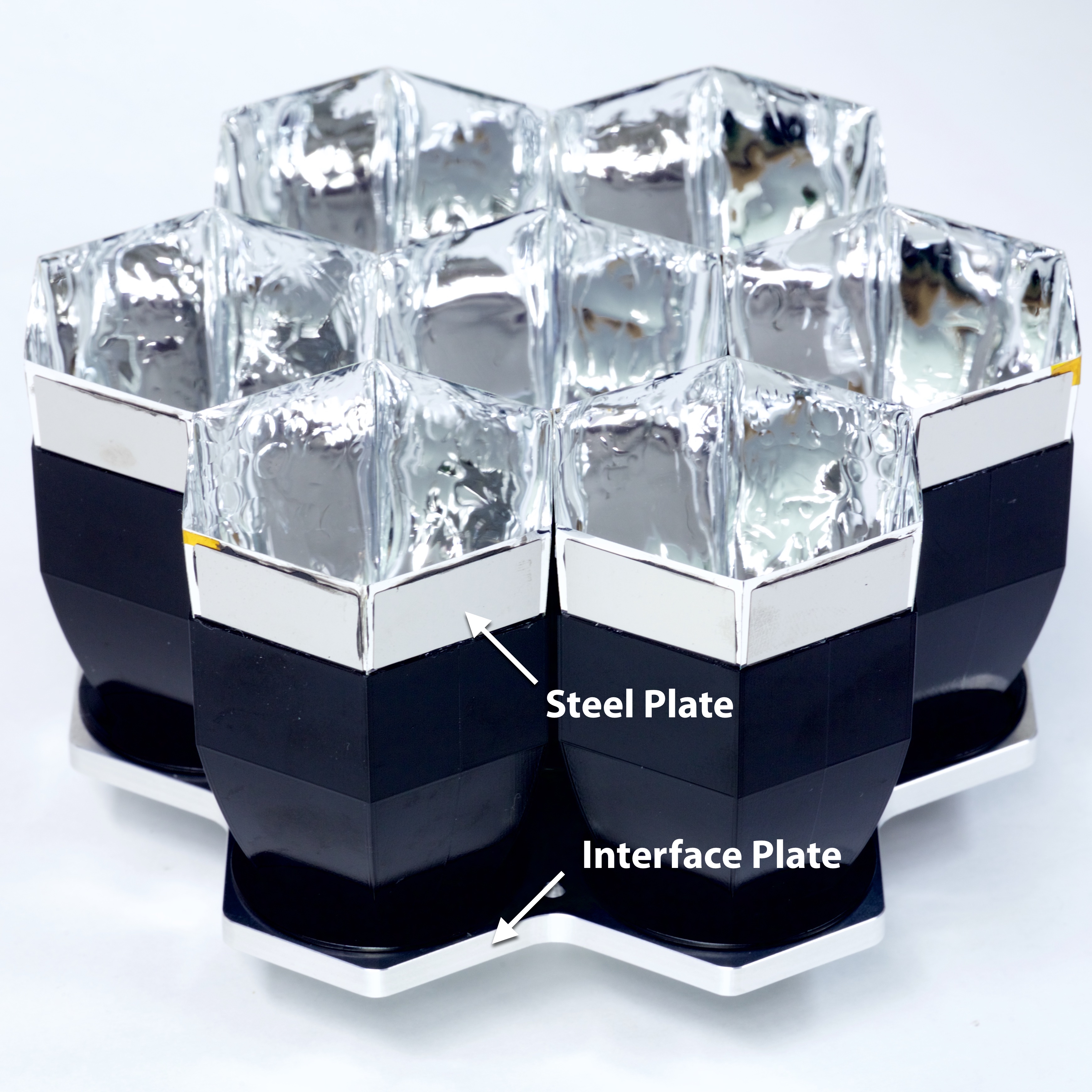}%
    \label{fig:7LCs_bird_view}
  }
\caption{(a) Seven light-concentrator prototypes viewed from above. The central prototype is dimmer than the others due to the ambient lighting and the camera position. (b) Same prototypes viewed from another direction. The thin steel plates and the interface plate indicated by arrows are visible from this direction. The steel plates are not in contact with the plastic cones.}
\label{fig:7LCs}
\end{figure}

Figure~\ref{fig:Glueing} shows the production process of light concentrators. First the six film pieces are aligned on a ``male'' block made of stainless steel, then silicone glue is put on the backside of each film piece as shown in Figure~\ref{fig:Glueing_with_hands}. The films and the male block are inserted into a plastic cone to tightly attach the films on the ABS surfaces (Figure~\ref{fig:Glueing_with_cable_binder}). A thin stainless steel plate (0.1~mm $\times$ 13~mm $\times$ 26~mm) is also glued onto the upper end of each film to keep it flat and solid (see Figure~\ref{fig:7LCs_bird_view}). It takes 20--30 minutes to complete these steps for a single cone.

Figure~\ref{fig:7LCs} shows a cluster of seven finished light concentrators attached to an interface plate (also see Figure~\ref{fig:CAD}). Owing to the thickness of ESR, it is apparent from Figure~\ref{fig:7LCs} that the dead area between neighboring pixels is negligible. The distortion of ESR pieces that is visible in Figure~\ref{fig:7LCs_bird_view} is caused by the thermal processes during multilayer coating in the vacuum chamber, but it does not significantly reduce the light-concentrator performance (see Figure~\ref{fig:simulation_465nm}) because it is not a precision-imaging device. In addition, very thin cracks are seen on the surface of the additional 54-layer coating, probably for the same reason. These cracks do not reduce the reflectance performance because their fraction of the specular area is negligible.

\section{Measurements and Results}
\label{sec:MeasRes}

\begin{figure}
\centering
\subfigure[]{%
  \includegraphics[width=.5715\textwidth,clip]{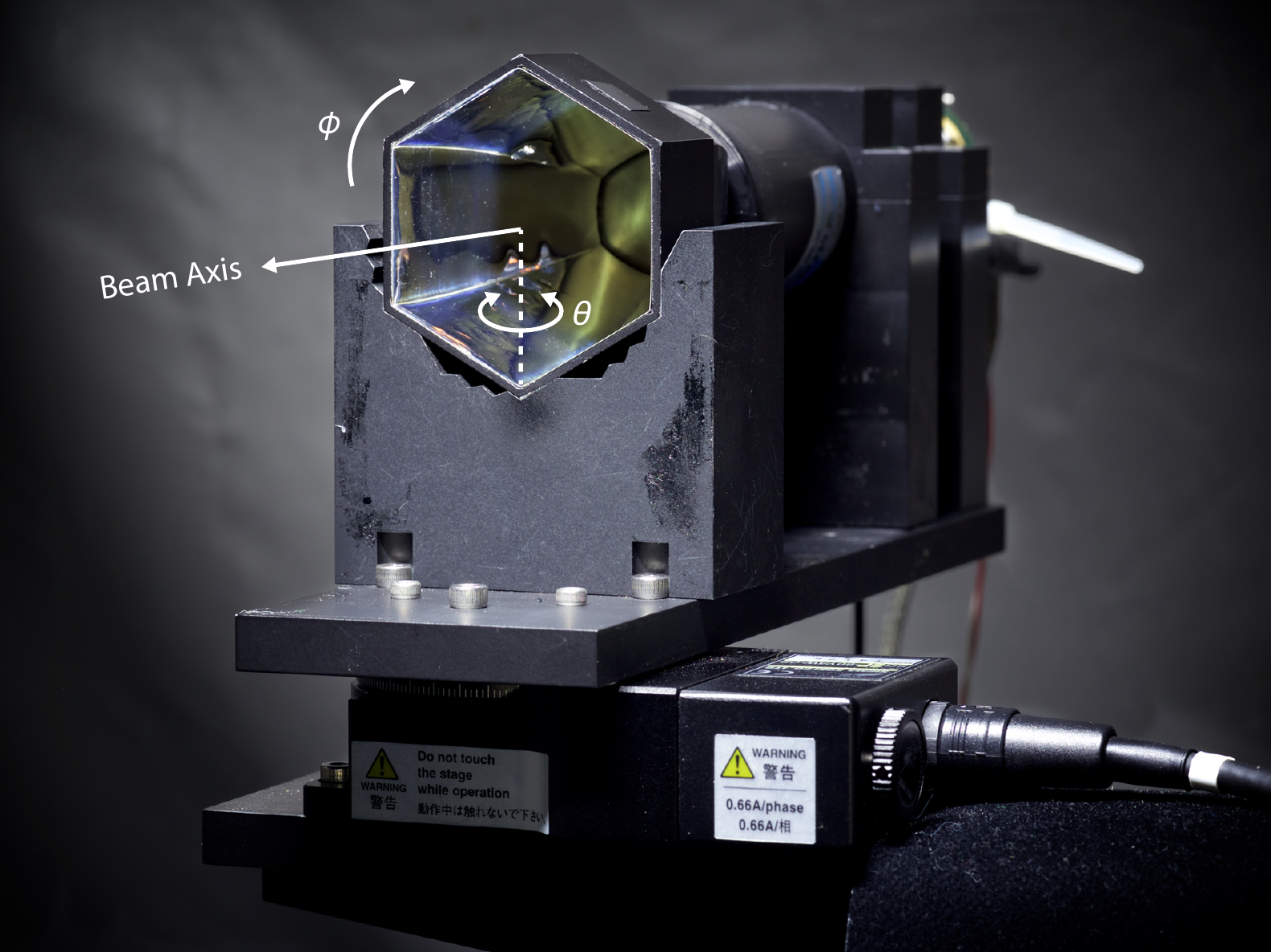}%
  \label{fig:Stage_with_cone}%
}%
\subfigure[]{%
  \includegraphics[width=.4285\textwidth,clip]{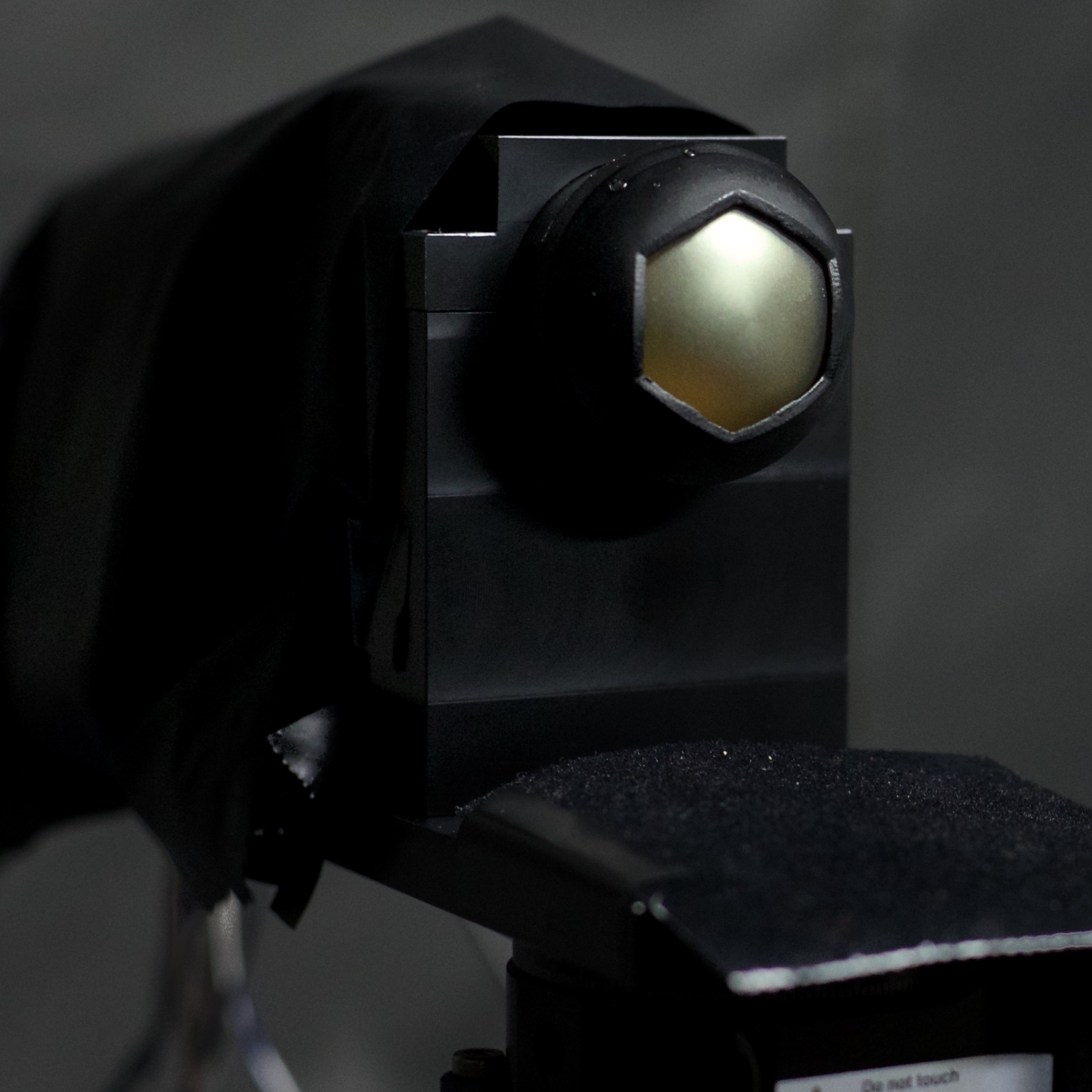}%
  \label{fig:Stage_with_mask}%
}
\caption{(a) Measurement system of a light concentrator in a dark room. The cone and a PMT module are placed on supporting blocks made of Delrin. They are attached to a rotation stage. The angle of incidence $\theta$ and rotation angle $\phi$ are defined as illustrated in the figure. This photo was taken with both $\theta$ and $\phi$ at $0^\circ$. (b) Photo of the 25-mm mask.}
\label{fig:Stage}
\end{figure}

The performances of the finished light concentrators were evaluated by measuring the relative anode sensitivity of an attached PMT for various angles of incidence, $\theta$, from $-40^\circ$ to $+40^\circ$ with $0.5^\circ$ ($20^\circ \le |\theta| \le 30^\circ$) or $1^\circ$ (otherwise) steps. The PMT and the light concentrator on a rotation stage were illuminated with blue or UV (465, 365, or 310~nm) LEDs, as shown in Figure~\ref{fig:Stage_with_cone}. The LEDs were placed behind a quartz diffuser whose distance to the entrance aperture of the light concentrator was 2.4~m. Therefore, the input light source is not an ideal parallel beam but a narrow cone with an opening angle of ${\sim}0.6^\circ$ ($=\tan^{-1}(25\ \mathrm{mm}/2400\ \mathrm{mm})$). The beam uniformity measured within a 70~mm~$\times$~70~mm region is about $\pm3$\%, where the PMT with the 5~mm-diameter hole was scanned in the square region by X and Z stages.

We define the relative anode sensitivity $Q$ as
\begin{equation}
  \label{eq:Q}
  Q = \frac{A_\mathrm{LC}(\theta)}{A_\mathrm{mask}} \times \frac{1}{3.7515\times\cos\theta}
\end{equation}
where $A_\mathrm{LC}(\theta)$ is the PMT-pulse area measured with a light concentrator and a waveform sampler, DRS4 \cite{Ritt:2008:Design-and-performance-of-the-6-GHz-waveform-digit}, at an angle of $\theta$, $A_\mathrm{mask}$ is the pulse area measured with a hexagon-like mask ($25\times\frac{2}{\sqrt{3}}$~mm vertex-to-vertex distance with slightly curved sides) at $\theta=0^\circ$ (see Figure~\ref{fig:Stage_with_mask}); and $3.7515$ is the ratio of the camera-pixel area (a 50-mm hexagon) to the mask\footnote{The mask area is slightly larger than a $25$-mm hexagon because of the hemispherical shape of the PMT; thus, $3.7515$ is used here instead of 4 (the area ratio of 25-mm and 50-mm hexagons).}. $Q$ can be regarded as the collection efficiency of the light concentrator if the angular and position dependence of the PMT-anode sensitivity (see Figure~\ref{fig:PMT_response}) are negligible. Note that it can exceed 100\%, because it is not efficiency by definition.

We have introduced the relative anode sensitivity $Q$ instead of the collection efficiency $\varepsilon$ for the following reason. The LST camera pixels are a coupled system of light concentrators and PMTs, in which the final photo-detection performance of the camera pixels is a convolution of the collection efficiency of a light concentrator and the position and angular dependent PMT sensitivity. Decoupling these factors is not useful for the CTA, and thus the relative anode sensitivity is used as our measure.

%We have introduced the relative anode sensitivity $Q$ for the following reasons: First it is difficult to measure a collection efficiency of a light concentrator because PMTs usually have angular and position dependence on their anode sensitivity. The distributions of photon entrance positions and angles of incidence are different between the presence or absence of the light concentrator; Thus what we can measure is always convolution of the collection efficiency and the non-constant anode sensitivity of the PMT, which should be evaluated not by collection efficiency but by relative anode sensitivity $Q$; The quality control and QE measurements of LST PMTs are performed without light concentrators. Therefore obtaining the final pixel performance as a relative anode sensitivity $Q$ is more important for calibration and Monte Carlo simulations.

The LEDs for the measurement were driven with an external function generator operated in burst mode. The PMT-pulse area was integrated for each trigger; the averaged pulse area of 500 triggers per angle is defined as $A_\mathrm{LC}(\theta)$.

\begin{landscape}
\begin{figure}
  \centering
  \subfigure[465~nm,  $\phi=0^\circ$]{%
    \includegraphics[width=.45\textwidth,clip]{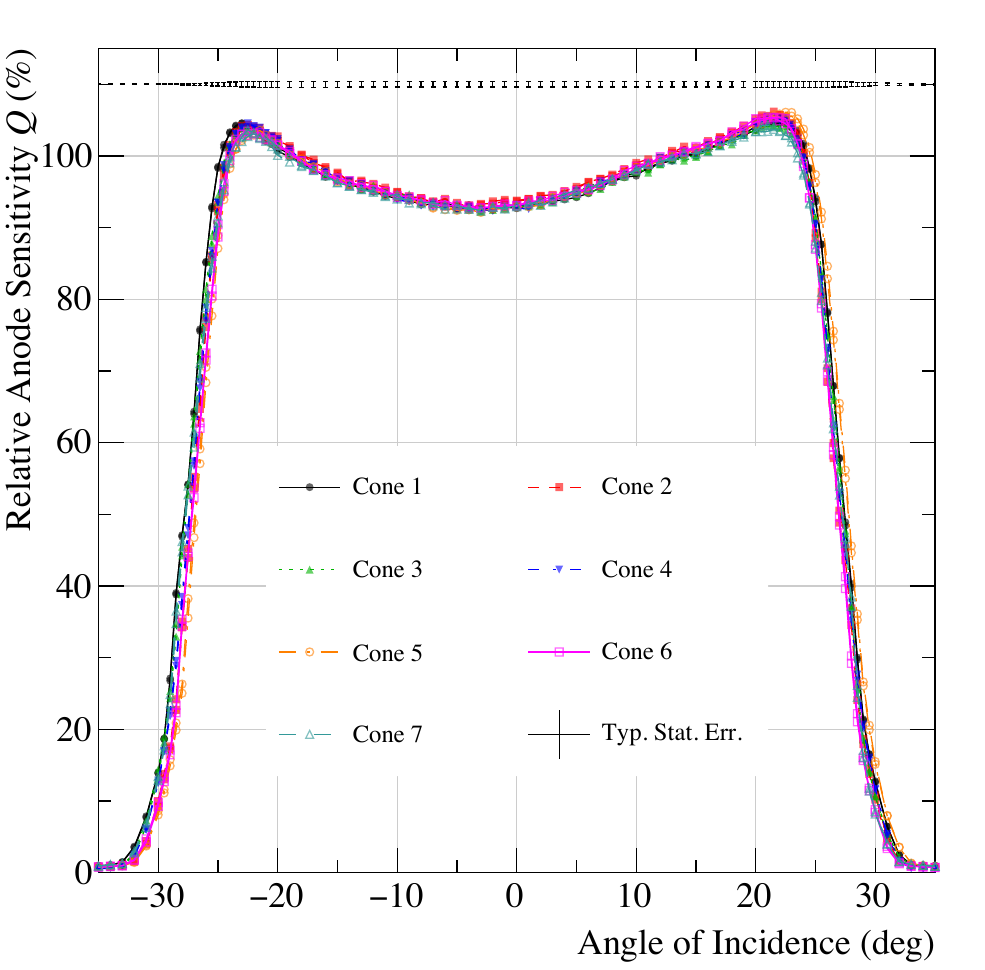}%
    \label{fig:465nm_0deg}
  }%
  \subfigure[365~nm,  $\phi=0^\circ$]{%
    \includegraphics[width=.45\textwidth,clip]{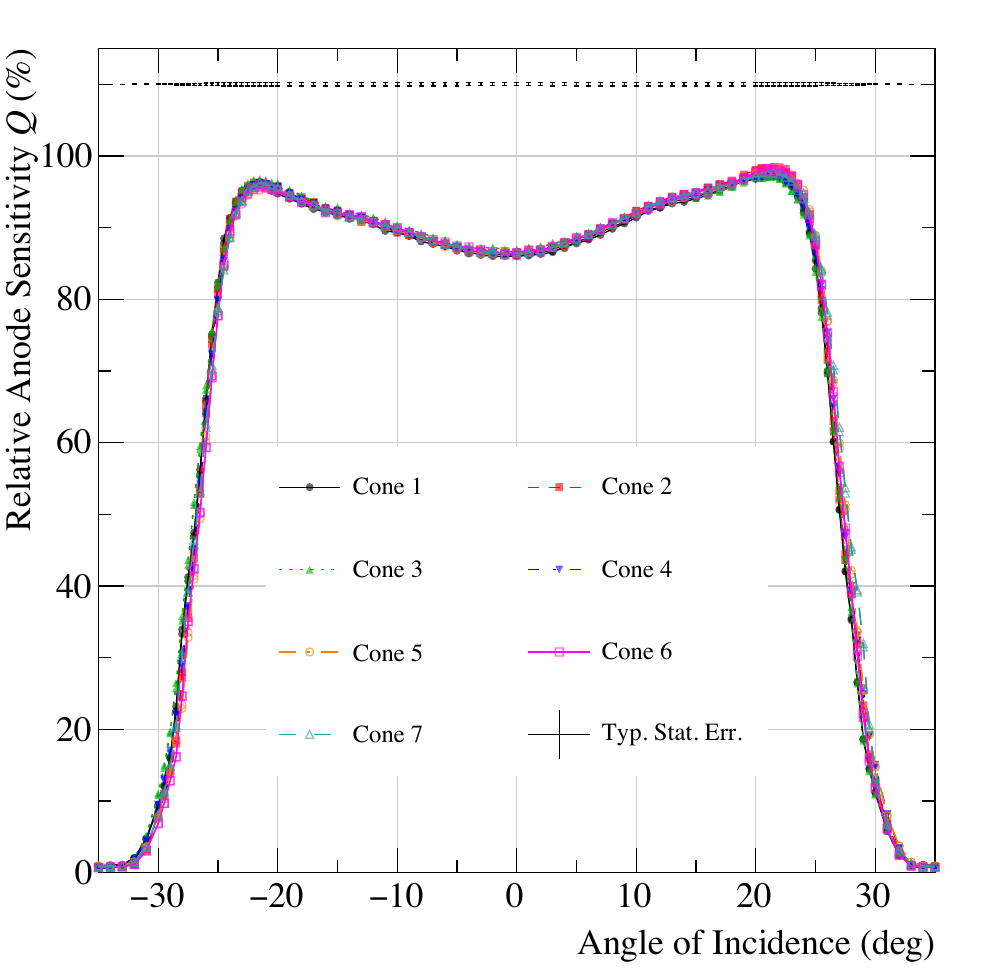}%
    \label{fig:365nm_0deg}
  }%
  \subfigure[310~nm,  $\phi=0^\circ$]{%
    \includegraphics[width=.45\textwidth,clip]{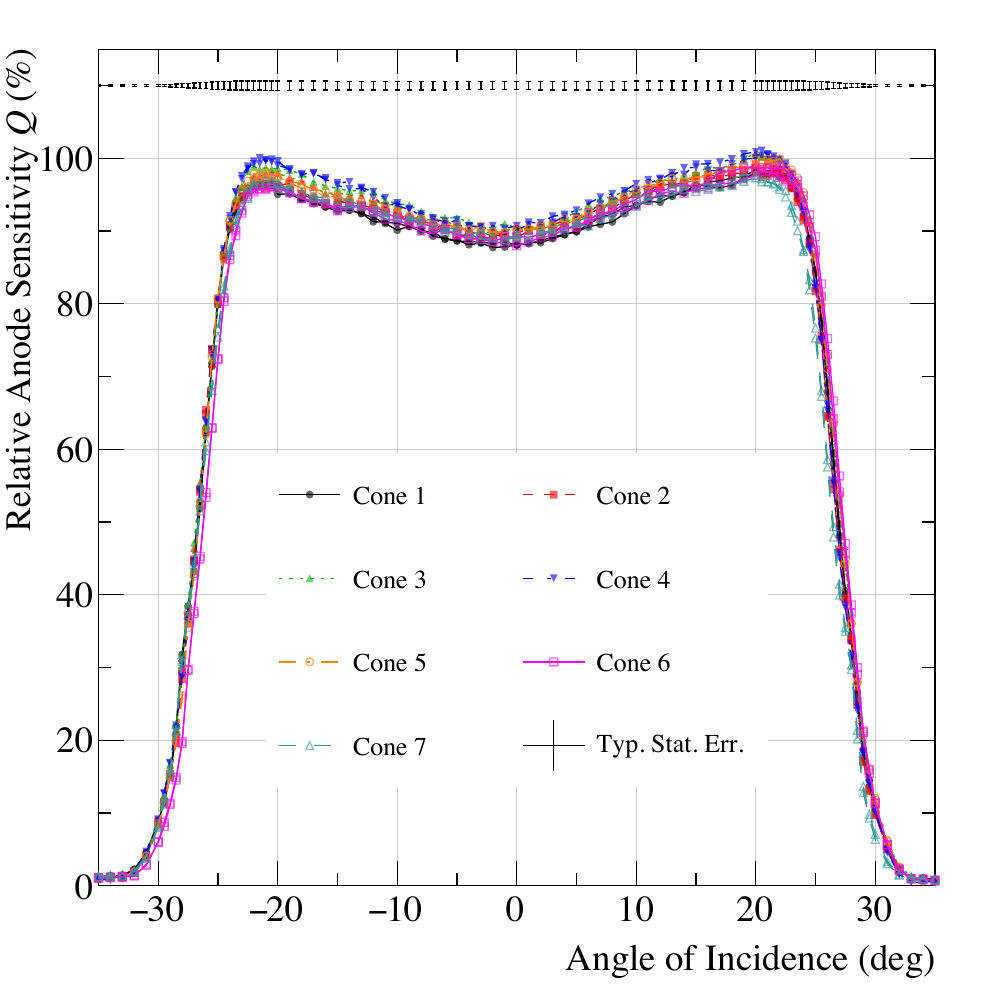}%
    \label{fig:310nm_0deg}
  }
  \subfigure[465~nm,  $\phi=30^\circ$]{%
    \includegraphics[width=.45\textwidth,clip]{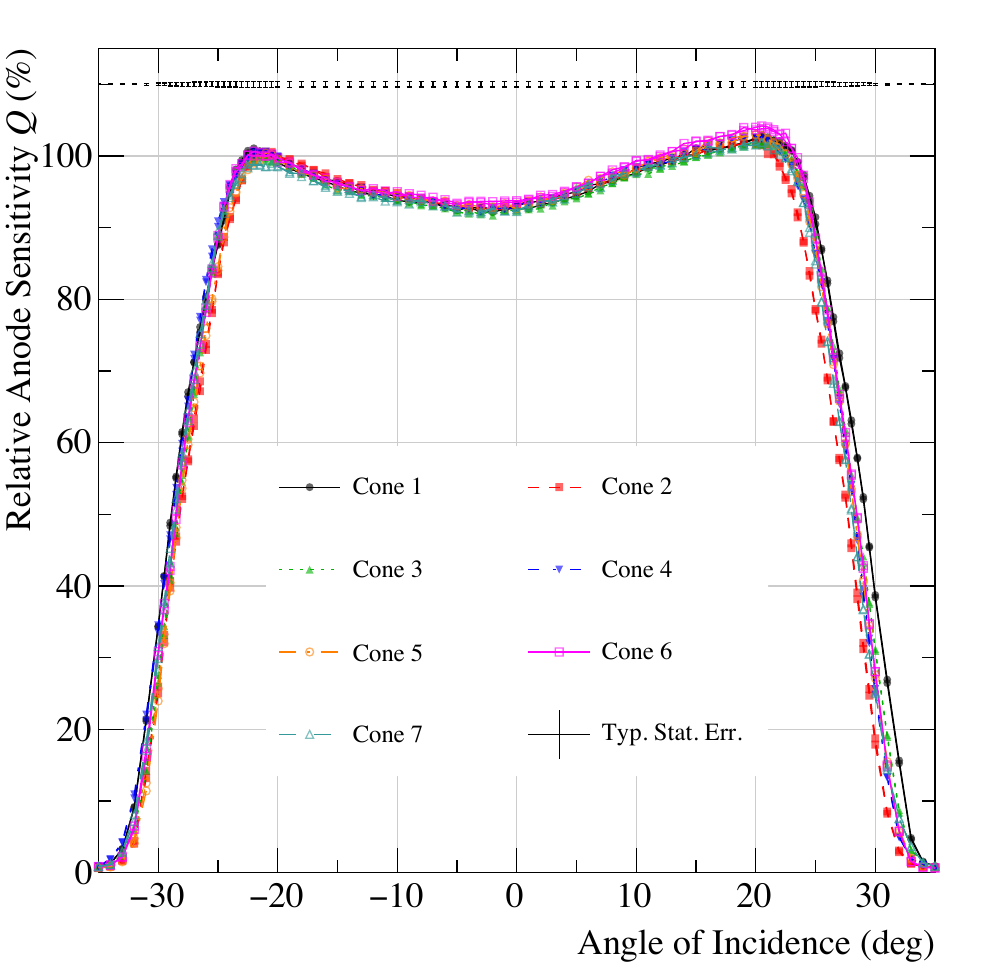}%
    \label{fig:465nm_30deg}
  }%
  \subfigure[365~nm,  $\phi=30^\circ$]{%
    \includegraphics[width=.45\textwidth,clip]{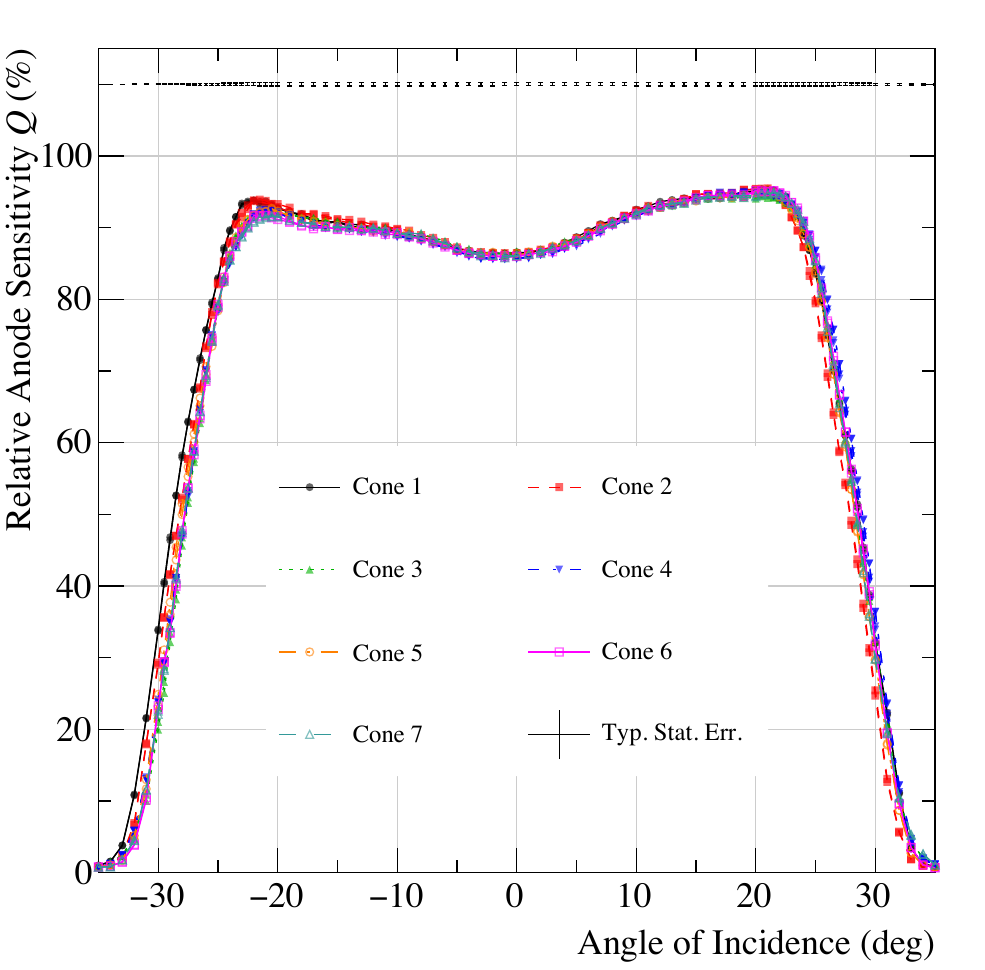}%
    \label{fig:365nm_30deg}
  }%
  \subfigure[310~nm,  $\phi=30^\circ$]{%
    \includegraphics[width=.45\textwidth,clip]{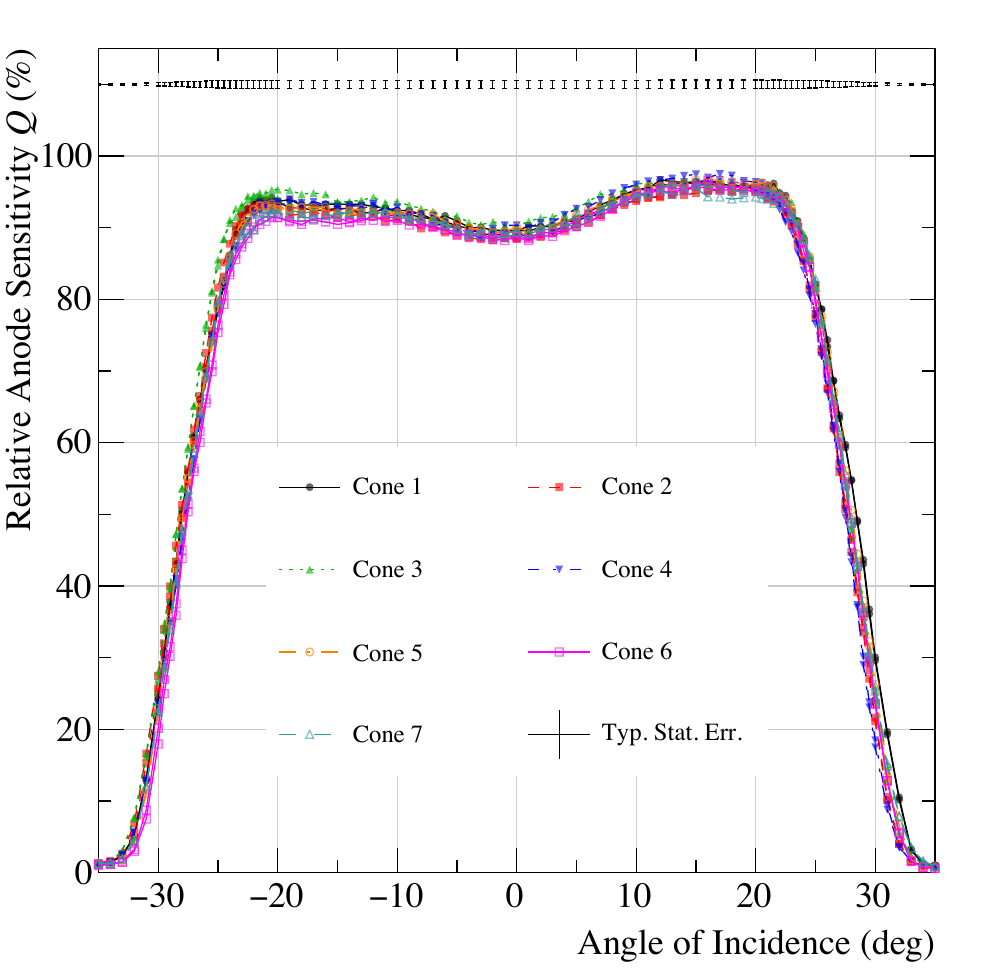}%
    \label{fig:310nm_30deg}
  }
\caption{(a) Measured relative anode sensitivities $Q$ for seven light concentrators as functions of the angle of incidence. The 465-nm LED was used with a rotation angle $\phi=0^\circ$. Two repeated measurements are drawn for each light concentrator; however, it is difficult to distinguish them because the difference is only about 0.5\%. Typical statistical errors of the data points are indicated with black error bars at $Y=110$\%. (b) 365~nm, $\phi=0^\circ$; (c) 310~nm,  $\phi=0^\circ$; (d) 465~nm,  $\phi=30^\circ$; (e) 365~nm,  $\phi=30^\circ$; and (f) 310~nm,  $\phi=30^\circ$.}
\label{fig:RAS}
\end{figure}
\end{landscape}

\begin{landscape}
\begin{figure}
  \centering
  \subfigure[465~nm,  $\phi=0^\circ$]{%
    \includegraphics[width=.45\textwidth,clip]{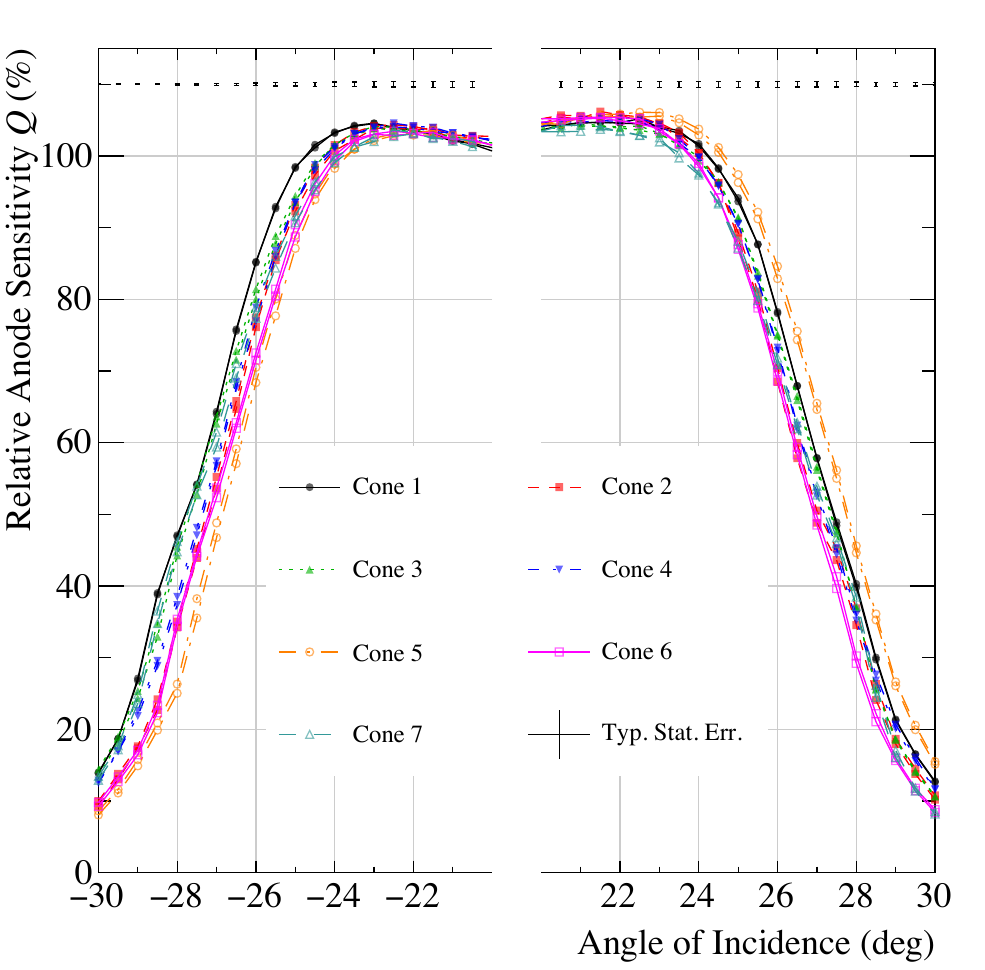}%
    \label{fig:465nm_0deg_zoom}
  }%
  \subfigure[365~nm,  $\phi=0^\circ$]{%
    \includegraphics[width=.45\textwidth,clip]{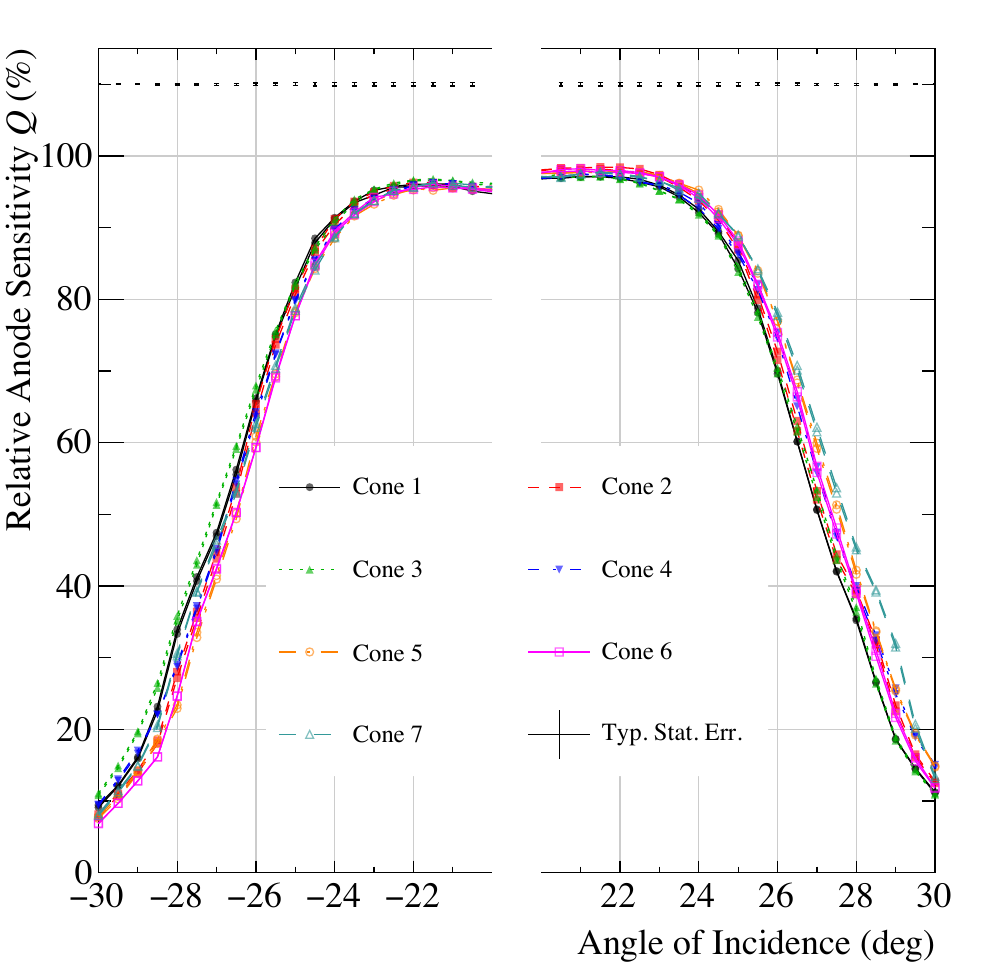}%
    \label{fig:365nm_0deg_zoom}
  }%
  \subfigure[310~nm,  $\phi=0^\circ$]{%
    \includegraphics[width=.45\textwidth,clip]{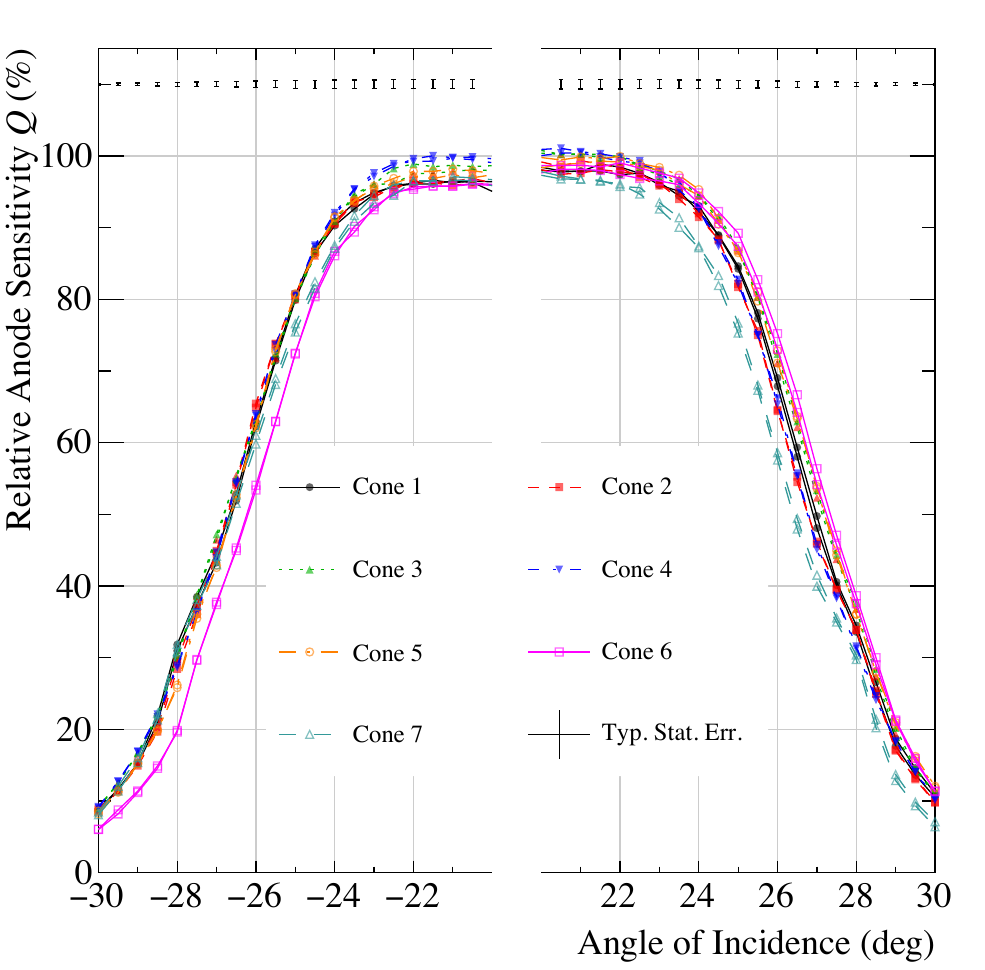}%
    \label{fig:310nm_0deg_zoom}
  }
  \subfigure[465~nm,  $\phi=30^\circ$]{%
    \includegraphics[width=.45\textwidth,clip]{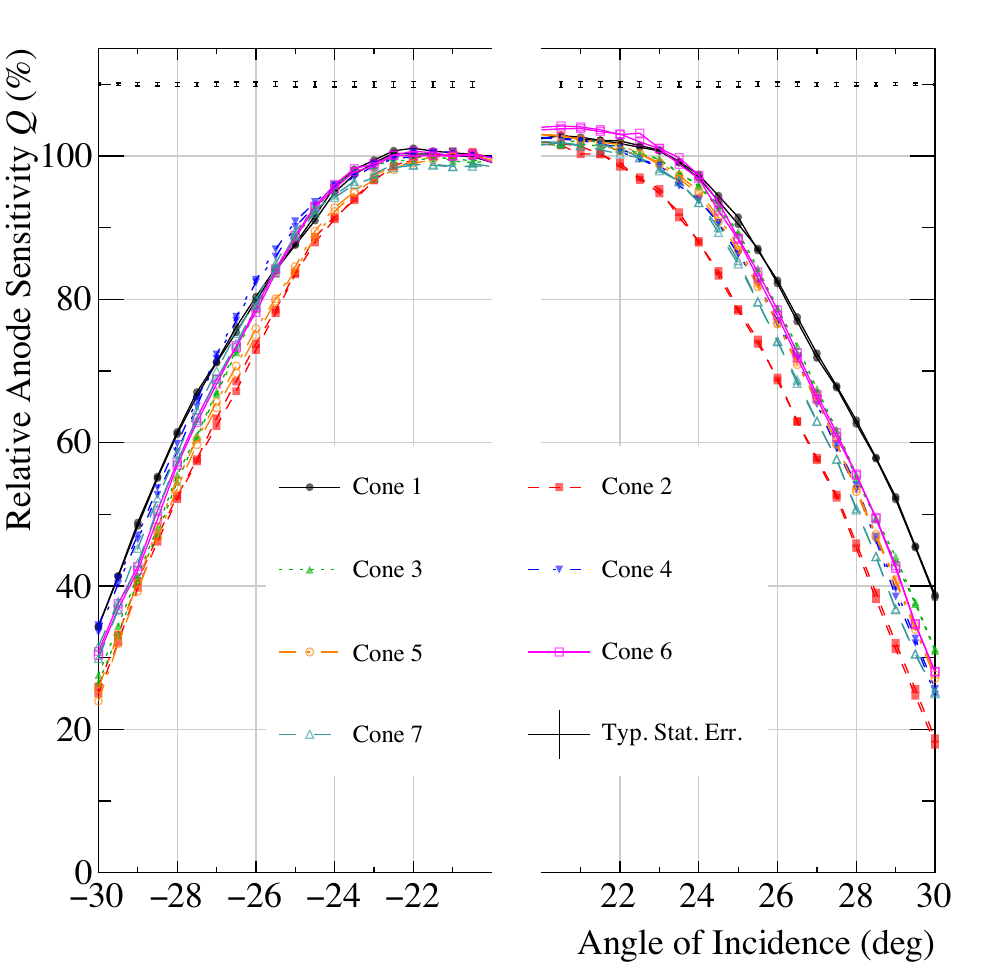}%
    \label{fig:465nm_30deg_zoom}
  }%
  \subfigure[365~nm,  $\phi=30^\circ$]{%
    \includegraphics[width=.45\textwidth,clip]{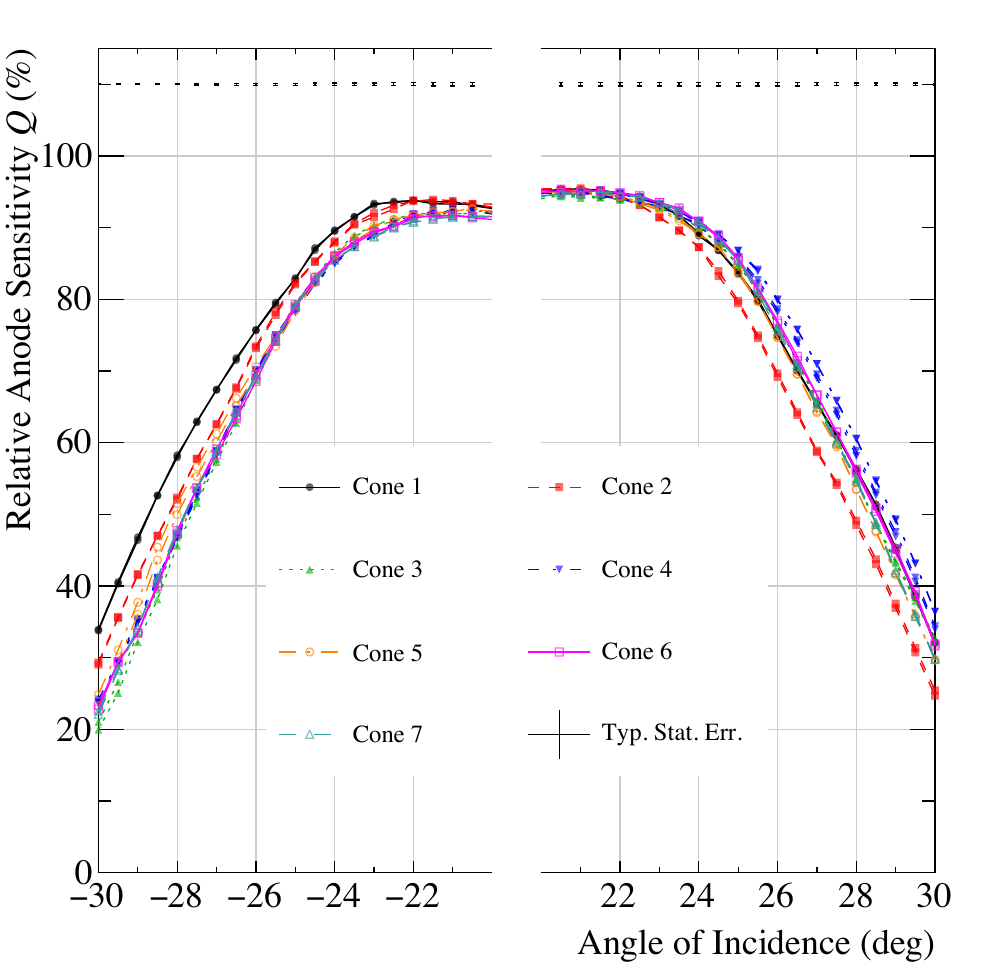}%
    \label{fig:365nm_30deg_zoom}
  }%
  \subfigure[310~nm,  $\phi=30^\circ$]{%
    \includegraphics[width=.45\textwidth,clip]{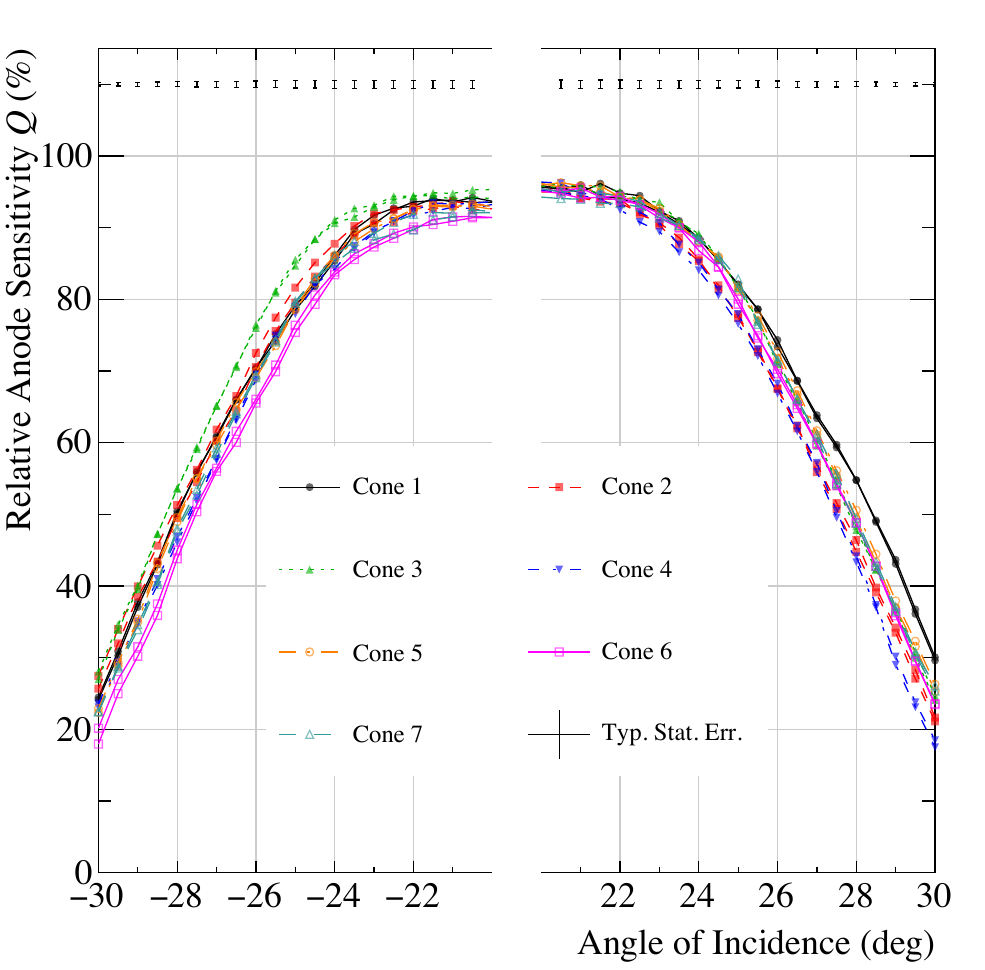}%
    \label{fig:310nm_30deg_zoom}
  }
\caption{Same as Figure~\ref{fig:RAS} but showing only limited angular regions (from $-30^\circ$ to $-20^\circ$ and from $20^\circ$ to $30^\circ$) to make individual curves more clearly visible.}
\label{fig:RAS_zoom}
\end{figure}
\end{landscape}

Figures~\ref{fig:RAS} and \ref{fig:RAS_zoom} show the relative anode sensitivity $Q$ versus the angle of incidence of seven light concentrators. Combinations of three LED colors and rotation angles, $\phi$, of $0^\circ$ and $30^\circ$ were measured (see Figure~\ref{fig:Stage_with_cone} for the definition of $\phi$). The main systematic uncertainties on this measurement, which affects the overall normalization, come from the scale factor $3.7515$ of the mask and the combined nonlinearity of the PMT, the amplifier, and the DRS4. The mask dimension has been checked by a micrometer, obtaining a systematic uncertainty of $0.3$\%. The nonlinearity in a good region is known to spread within about $\pm1$\% \cite{Masuda:2015:Development-of-the-photomultiplier-tube-readout-sy}, and thus the total systematic uncertainty is derived to be typically about $1$\%.

As shown in Figure~\ref{fig:465nm_0deg}, for example, the relative anode sensitivity $Q$ reaches the highest value at around $\theta=23^\circ$. This is because the hemispherical photocathode of R11920 has a higher anode sensitivity for larger angles of incidence, as shown in Figure~\ref{fig:PMT_angle}.

The overall shapes of the relative anode sensitivity $Q$ of the seven cones are asymmetrical around $\theta=0^\circ$, but they are very similar to each other. This means that the asymmetry is not due to the shape of the produced light concentrators but is rather due the position and angular dependence of the PMT photocathode.

The uniformity of the relative anode sensitivity $Q$ of the seven cones is within ${\sim}2$\% in $-23^\circ\le\theta\le23^\circ$ for 465~nm and 365~nm, which is sufficiently small but a few times larger than the systematic error of the normalization (${\sim}1$\%), the typical statistical error of 0.4\%, or the repeat accuracy of the measurement of ${\sim}0.5$\%. On the contrary, the relative anode sensitivity $Q$ measured with the 310-nm LED varies by ${\sim}5$\%, as can be seen from Figures~\ref{fig:310nm_0deg}, \ref{fig:310nm_30deg}, \ref{fig:310nm_0deg_zoom}, and \ref{fig:310nm_30deg_zoom}. This is presumably because the reflectance of UV-enhanced ESR films around 310~nm is not very stable due to the difficulty in precision control of the thickness of the multilayer coating; thus, its final performance as a light concentrator varies film by film.

The cutoff profile around $\theta=27^\circ$ is stable within about $1^\circ$, including an alignment error of about $0.5^\circ$ during the measurements. This cutoff variation is not a significant problem because the fraction of photons with $\theta\ge25^\circ$ is small (see Figure~\ref{fig:AngDist}).

\begin{landscape}
\begin{figure}
  \centering
  \subfigure[465~nm]{%
    \includegraphics[width=.5\textwidth,clip]{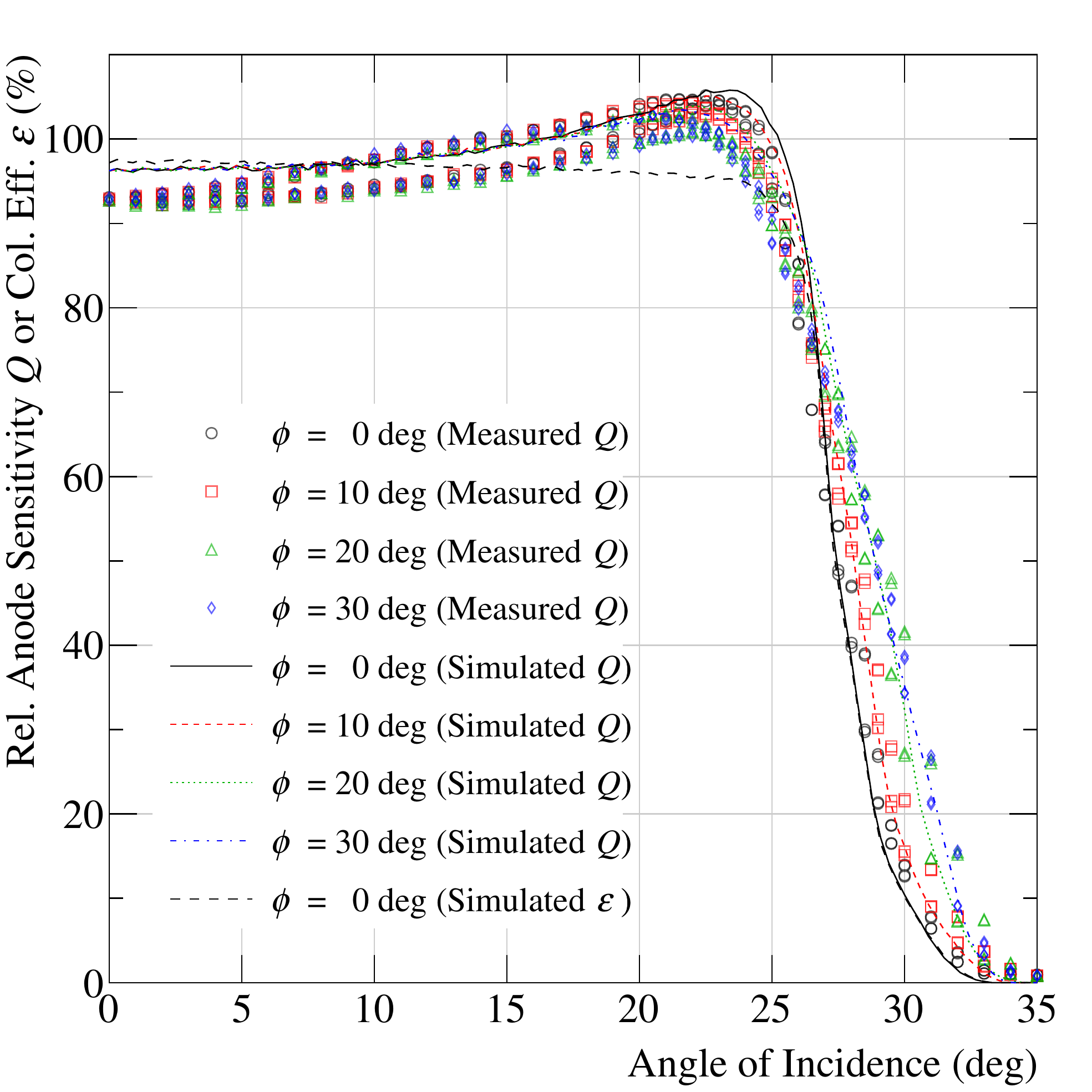}%
    \label{fig:simulation_465nm}
  }%
  \subfigure[365~nm]{%
    \includegraphics[width=.5\textwidth,clip]{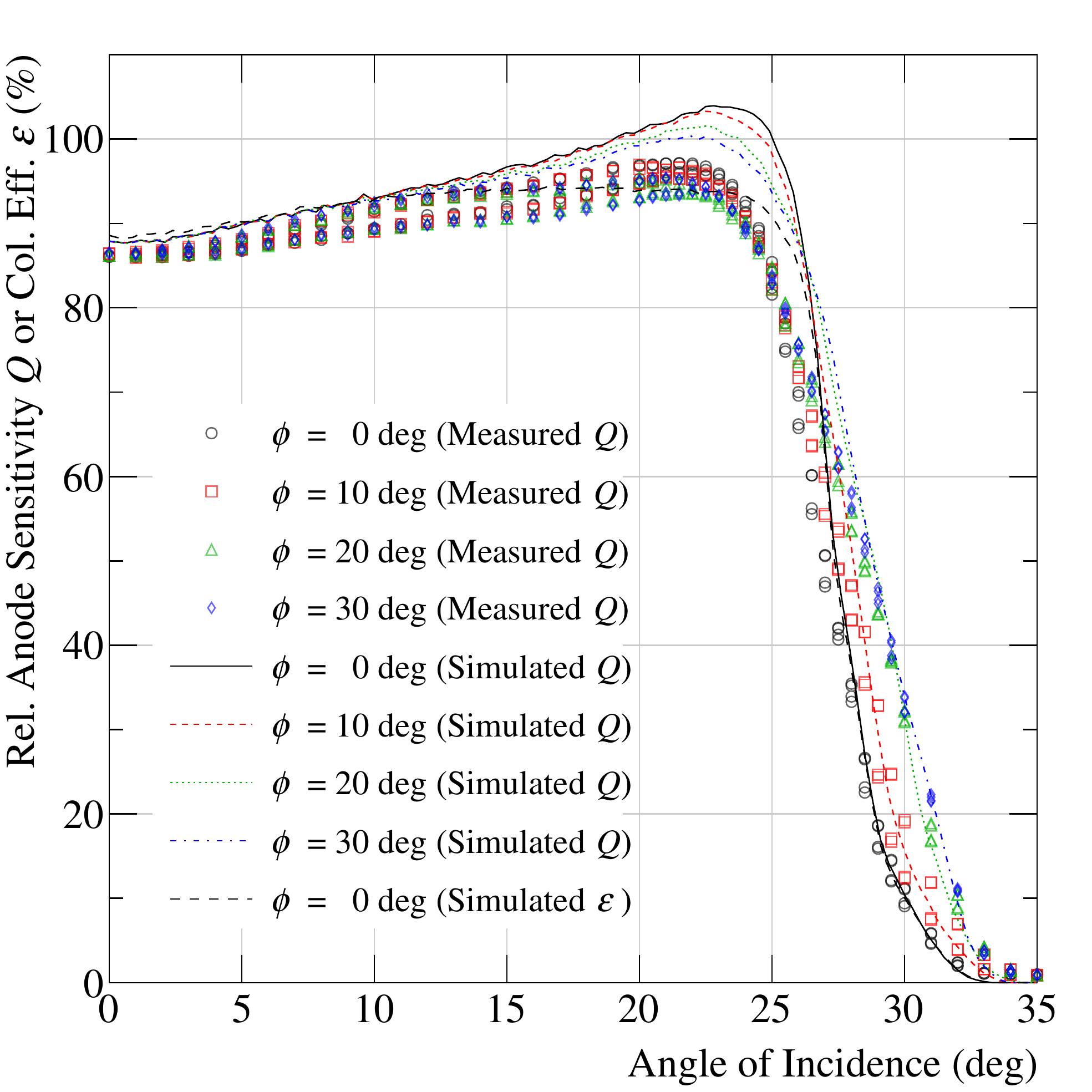}%
    \label{fig:simulation_365nm}
  }%
  \subfigure[310~nm]{%
    \includegraphics[width=.5\textwidth,clip]{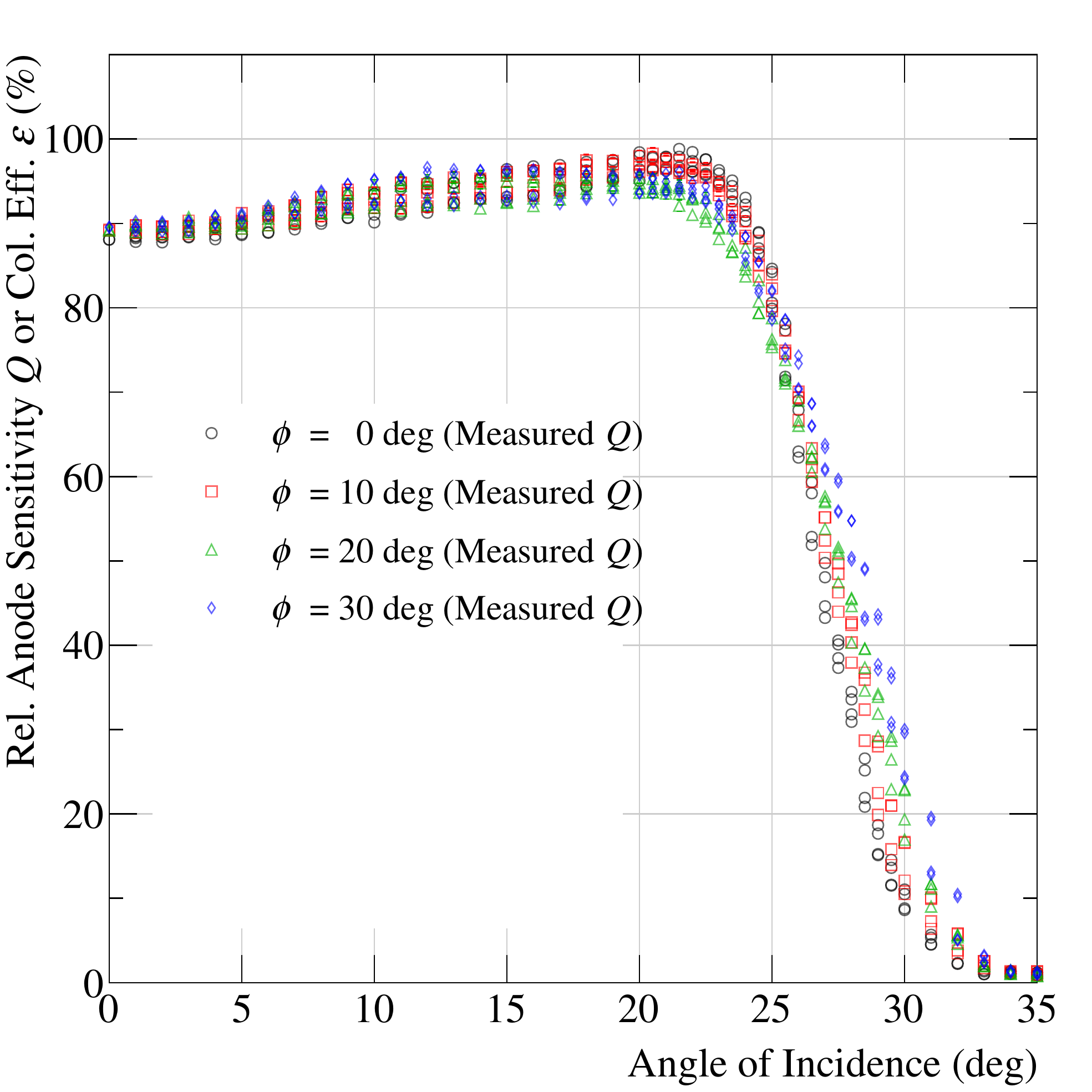}%
    \label{fig:simulation_310nm}
  }%
\caption{(a) Comparison of simulated and measured relative anode sensitivities $Q$ for a wavelength of 465~nm. Measurements of Cone~1 (see Figures~\ref{fig:RAS} and \ref{fig:RAS_zoom}) with rotation angles, $\phi$, of $0$, $10^\circ$, $20^\circ$, and $30^\circ$ are shown. The data points for negative angles of incidence in Figures~\ref{fig:RAS} are also overlaid by inverting the sign. The long dash line shows a simulated cone collection efficiency $\varepsilon$ for $\phi=0^{\circ}$. (b) Same as (a) but for a wavelength of 365~nm. (c) Same as (a) but for a wavelength of 310~nm and without simulation.}
\label{fig:simulation}
\end{figure}
\end{landscape}

Figure~\ref{fig:simulation} compares the measured relative anode sensitivity $Q$ with a \texttt{ROBAST} simulation in which the averaged angular and positional dependence of PMTs' anode sensitivity (see Figures~\ref{fig:PMT_angle} and \ref{fig:PMT_position}) as well as the angular and wavelength dependence of the reflectance of UV-enhanced ESR (see Figure~\ref{fig:reflectance}) were taken into account. We could measure the reflectance only with a limited number of angles from $20^\circ$ up to $70^\circ$ with a step size of $10^\circ$ owing to the mechanical limitations of the spectrometer. Thus, we applied spline interpolation and extrapolation of the reflectance by searching and connecting the peak structures shown in Figure~\ref{fig:reflectance_zoomin}, so as to cover all angles of incidence in the simulation\footnote{Spectrum ``blue shift'' appears in the multilayer coating when the angle of incidence becomes large. For example, the most noticeable blue shift in Figure~\ref{fig:reflectance_zoomin} can be found when a spectrum peak at ${\sim}600$~nm ($20^\circ$) shifts to ${\sim}500$~nm ($70^\circ$). This shift is due to the fact that the effective layer thickness increases as the angle of incidence increases. }. Because of the instability of the spectrometer below 320~nm, we could not apply interpolation or extrapolation for 310~nm; thus, \texttt{ROBAST} simulation could not be performed in Figure~\ref{fig:simulation_310nm}.

 The collection efficiency\footnote{We define the collection efficiency $\varepsilon$ to be the ratio between the number of photons illuminated on a 50-mm hexagon to the number of photons exiting from the cone.} $\varepsilon$ of the cone was also simulated without a PMT and is compared with $Q$ in Figure~\ref{fig:simulation}. The simulated $\varepsilon$ does not have the shoulder structure seen in $Q$ around $23^{\circ}$.

The simulation and measurement results of $Q$ are generally consistent with each other to within about 5\% for 465~nm up to 25$^\circ$. The peak structure around $23^\circ$ is well reproduced in the simulation due to the assumption of angular dependence of the PMT sensitivity (Figure~\ref{fig:PMT_angle}), without which the simulated collection efficiency $\varepsilon$ curve becomes flatter from $0^\circ$ to $25^\circ$.

The first significant discrepancy between the simulation and the measurement is that the peak position of the measured data at both 465 and 365~nm is shifted by about $-1^\circ$ from the simulation. This is probably because the upper edge of each ESR film is supported only by a smaller thin steel plate; therefore, the actual cone shape is slightly different from the ideal profile assumed in the simulation, resulting in photon reflections to unexpected directions.

The second discrepancy is at the maximum of 365~nm. The measured sensitivity is 5--20\% lower than the simulation in the angular range from $20^\circ$ to $27^\circ$. One plausible reason is that the angular dependence shown in Figure~\ref{fig:PMT_angle}, which was measured with the 465-nm LED only at the photocathode center, may be dependent upon wavelength or position. To understand this discrepancy, further study of the photocathode sensitivity using dedicated measurements is needed. Another reason is that the assumed interpolation and extrapolation of the reflectance for 365~nm assumed in the simulation may not be accurate. This is because the reflectance change between $60^\circ$ and $70^\circ$ is rapid around 365~nm (Figure~\ref{fig:reflectance_zoomin}).

In a previous study, the light collection efficiency of a hexagonal light concentrator with an entrance aperture of 23.2~mm was measured for SST-1M, which is a single composite-mirror telescope design (a Davies--Cotton optical system) proposed for the CTA Southern array \cite{Aguilar:2015:Design-optimization-and-characterization-of-the-li}. The cone is produced by injection molding to have a cutoff angle of 24$^\circ$, the inner reflective surfaces of which have aluminum and reflection enhancement coating for UV-blue photons. The measured light collection efficiencies range between about 88--92\%, 84--92\%, and 80--92\%, for 355~nm, 390~nm, and 595~nm, respectively, for the angles of incidences between 10--20$^\circ$. Taking into account its plastic thickness of 0.5 mm, the net collection efficiency is obtained by multiplying 0.919 ($=23.2^2/24.2^2$) to these values  (see Fig.~\ref{fig:simulation} for our simulated collection efficiency).

Prototype light concentrators for NectarCAM, one of two camera designs of CTA MSTs, were also produced by using similar techniques used by the SST-1M team, and their performance was measured by comparing the PMT anode current with a light concentrator and with a 23.4-mm hexagonal mask \cite{Henault:2017:Testing-light-concentrators-prototypes-for-the-Che} as we did with the 25-mm hexagon-like mask. Their cone performance measured with the same PMT series (Hamamatsu Photonics R11920-100) as ours and six different colors (325, 390, 420, 440, 480, and 517~nm) was characterized as ``rejection curves,'' in which the factor $1/\cos\theta$ used in Equation~(\ref{eq:Q}) is not multiplied for their normalization. In addition, they used a scaling factor of $4.39$ instead of our $3.7515$ in Equation~(\ref{eq:Q}) because of the mask size difference and because they normalized the rejection curves for a 49-mm aperture, while 50 mm for ours. Taking into account these differences, our relative anode sensitivities are roughly 5--10\% better then theirs.

These comparison between the LST, MST, and SST-1M light concentrator prototypes shows the advantage of our design in photon detection, while more accurate comparison of the different techniques of light concentrator production would require the use of the same test bench and the cone geometry.

\section{Conclusion}
\label{sec:Conclusion}

We have developed a prototype hexagonal light concentrator by placing very reflective specular films on an ABS plastic cone for use in the Large-Sized Telescopes of the Cherenkov Telescope Array (CTA). Its relative anode sensitivity $Q$, which can be interpreted as the collection efficiency in the first approximation, was measured for seven cones with three LED colors for various angles of incidence and rotation angles. The sensitivity curves are consistent with simulation, and the maximum relative anode sensitivity $Q$ reaches ${\sim}95$ to ${\sim}105$\%.

\acknowledgments

We are grateful to the Industrial Technology Institute of Ibaraki Prefecture, Japan, where the reflectance of normal and UV-enhanced ESR films was measured. We also gratefully acknowledge the Technical Center of Nagoya University, where a number of early prototypes were produced by using a 3D printer. We also thank a number of colleagues in the CTA Consortium who helped us develop the light concentrators. In particular, Dr. Eckart Lorenz firstly brought the idea of additional multilayer coating on normal ESR. This study was supported by JSPS KAKENHI Grant Number 25707017 and Max-Planck-Institut f\"{u}r Kernphysik (MPIK). A.~O. was supported by a Grant-in-Aid for JSPS Fellows and MPIK when he worked as a guest scientist at the University of Leicester, UK and MPIK, respectively. A part of the ray-tracing simulation in this work was performed using computer resources at the Institute for Cosmic Ray Research (ICRR), the University of Tokyo. Finally we thank the anonymous journal referee for their helpful comments which have improved the paper quality.\\\\This paper has gone through internal review by the CTA Consortium.

\end{document}